\documentclass[aps,pra,reprint,showpacs,nofootinbib,superscriptaddress]{revtex4-1}
\usepackage{graphicx, comment}
\usepackage{amsfonts}
\usepackage{amsmath,amssymb}
\usepackage{bm}
\usepackage{color}
\usepackage{epstopdf}
\usepackage{epsfig}
\usepackage{verbatim}
\usepackage{lineno}

\usepackage{subcaption}
\usepackage{ragged2e}
\DeclareCaptionJustification{justified}{\justifying}
\usepackage[thicklines]{cancel}
\usepackage{url}

\usepackage{todonotes}
\usepackage{enumitem}
\usepackage{xcolor}
\usepackage{listings}
\usepackage{algorithm,algorithmic}
\usepackage{enumerate}
\usepackage{siunitx}
\usepackage{booktabs}
\usepackage{hyperref}
\hypersetup{
     colorlinks   = true,
     citecolor    = blue
}
\definecolor{red}{rgb}{1,0.,0}

\usepackage[compatibility=false]{caption}

\usepackage[draft]{minted}
\usepackage{tcolorbox}

\makeatletter
\def\PYG@reset{\let\PYG@it=\relax \let\PYG@bf=\relax%
    \let\PYG@ul=\relax \let\PYG@tc=\relax%
    \let\PYG@bc=\relax \let\PYG@ff=\relax}
\def\PYG@tok#1{\csname PYG@tok@#1\endcsname}
\def\PYG@toks#1+{\ifx\relax#1\empty\else%
    \PYG@tok{#1}\expandafter\PYG@toks\fi}
\def\PYG@do#1{\PYG@bc{\PYG@tc{\PYG@ul{%
    \PYG@it{\PYG@bf{\PYG@ff{#1}}}}}}}
\def\PYG#1#2{\PYG@reset\PYG@toks#1+\relax+\PYG@do{#2}}

\expandafter\def\csname PYG@tok@w\endcsname{\def\PYG@tc##1{\textcolor[rgb]{0.73,0.73,0.73}{##1}}}
\expandafter\def\csname PYG@tok@c\endcsname{\let\PYG@it=\textit\def\PYG@tc##1{\textcolor[rgb]{0.25,0.50,0.50}{##1}}}
\expandafter\def\csname PYG@tok@cp\endcsname{\def\PYG@tc##1{\textcolor[rgb]{0.74,0.48,0.00}{##1}}}
\expandafter\def\csname PYG@tok@k\endcsname{\let\PYG@bf=\textbf\def\PYG@tc##1{\textcolor[rgb]{0.00,0.50,0.00}{##1}}}
\expandafter\def\csname PYG@tok@kp\endcsname{\def\PYG@tc##1{\textcolor[rgb]{0.00,0.50,0.00}{##1}}}
\expandafter\def\csname PYG@tok@kt\endcsname{\def\PYG@tc##1{\textcolor[rgb]{0.69,0.00,0.25}{##1}}}
\expandafter\def\csname PYG@tok@o\endcsname{\def\PYG@tc##1{\textcolor[rgb]{0.40,0.40,0.40}{##1}}}
\expandafter\def\csname PYG@tok@ow\endcsname{\let\PYG@bf=\textbf\def\PYG@tc##1{\textcolor[rgb]{0.67,0.13,1.00}{##1}}}
\expandafter\def\csname PYG@tok@nb\endcsname{\def\PYG@tc##1{\textcolor[rgb]{0.00,0.50,0.00}{##1}}}
\expandafter\def\csname PYG@tok@nf\endcsname{\def\PYG@tc##1{\textcolor[rgb]{0.00,0.00,1.00}{##1}}}
\expandafter\def\csname PYG@tok@nc\endcsname{\let\PYG@bf=\textbf\def\PYG@tc##1{\textcolor[rgb]{0.00,0.00,1.00}{##1}}}
\expandafter\def\csname PYG@tok@nn\endcsname{\let\PYG@bf=\textbf\def\PYG@tc##1{\textcolor[rgb]{0.00,0.00,1.00}{##1}}}
\expandafter\def\csname PYG@tok@ne\endcsname{\let\PYG@bf=\textbf\def\PYG@tc##1{\textcolor[rgb]{0.82,0.25,0.23}{##1}}}
\expandafter\def\csname PYG@tok@nv\endcsname{\def\PYG@tc##1{\textcolor[rgb]{0.10,0.09,0.49}{##1}}}
\expandafter\def\csname PYG@tok@no\endcsname{\def\PYG@tc##1{\textcolor[rgb]{0.53,0.00,0.00}{##1}}}
\expandafter\def\csname PYG@tok@nl\endcsname{\def\PYG@tc##1{\textcolor[rgb]{0.63,0.63,0.00}{##1}}}
\expandafter\def\csname PYG@tok@ni\endcsname{\let\PYG@bf=\textbf\def\PYG@tc##1{\textcolor[rgb]{0.60,0.60,0.60}{##1}}}
\expandafter\def\csname PYG@tok@na\endcsname{\def\PYG@tc##1{\textcolor[rgb]{0.49,0.56,0.16}{##1}}}
\expandafter\def\csname PYG@tok@nt\endcsname{\let\PYG@bf=\textbf\def\PYG@tc##1{\textcolor[rgb]{0.00,0.50,0.00}{##1}}}
\expandafter\def\csname PYG@tok@nd\endcsname{\def\PYG@tc##1{\textcolor[rgb]{0.67,0.13,1.00}{##1}}}
\expandafter\def\csname PYG@tok@s\endcsname{\def\PYG@tc##1{\textcolor[rgb]{0.73,0.13,0.13}{##1}}}
\expandafter\def\csname PYG@tok@sd\endcsname{\let\PYG@it=\textit\def\PYG@tc##1{\textcolor[rgb]{0.73,0.13,0.13}{##1}}}
\expandafter\def\csname PYG@tok@si\endcsname{\let\PYG@bf=\textbf\def\PYG@tc##1{\textcolor[rgb]{0.73,0.40,0.53}{##1}}}
\expandafter\def\csname PYG@tok@se\endcsname{\let\PYG@bf=\textbf\def\PYG@tc##1{\textcolor[rgb]{0.73,0.40,0.13}{##1}}}
\expandafter\def\csname PYG@tok@sr\endcsname{\def\PYG@tc##1{\textcolor[rgb]{0.73,0.40,0.53}{##1}}}
\expandafter\def\csname PYG@tok@ss\endcsname{\def\PYG@tc##1{\textcolor[rgb]{0.10,0.09,0.49}{##1}}}
\expandafter\def\csname PYG@tok@sx\endcsname{\def\PYG@tc##1{\textcolor[rgb]{0.00,0.50,0.00}{##1}}}
\expandafter\def\csname PYG@tok@m\endcsname{\def\PYG@tc##1{\textcolor[rgb]{0.40,0.40,0.40}{##1}}}
\expandafter\def\csname PYG@tok@gh\endcsname{\let\PYG@bf=\textbf\def\PYG@tc##1{\textcolor[rgb]{0.00,0.00,0.50}{##1}}}
\expandafter\def\csname PYG@tok@gu\endcsname{\let\PYG@bf=\textbf\def\PYG@tc##1{\textcolor[rgb]{0.50,0.00,0.50}{##1}}}
\expandafter\def\csname PYG@tok@gd\endcsname{\def\PYG@tc##1{\textcolor[rgb]{0.63,0.00,0.00}{##1}}}
\expandafter\def\csname PYG@tok@gi\endcsname{\def\PYG@tc##1{\textcolor[rgb]{0.00,0.63,0.00}{##1}}}
\expandafter\def\csname PYG@tok@gr\endcsname{\def\PYG@tc##1{\textcolor[rgb]{1.00,0.00,0.00}{##1}}}
\expandafter\def\csname PYG@tok@ge\endcsname{\let\PYG@it=\textit}
\expandafter\def\csname PYG@tok@gs\endcsname{\let\PYG@bf=\textbf}
\expandafter\def\csname PYG@tok@gp\endcsname{\let\PYG@bf=\textbf\def\PYG@tc##1{\textcolor[rgb]{0.00,0.00,0.50}{##1}}}
\expandafter\def\csname PYG@tok@go\endcsname{\def\PYG@tc##1{\textcolor[rgb]{0.53,0.53,0.53}{##1}}}
\expandafter\def\csname PYG@tok@gt\endcsname{\def\PYG@tc##1{\textcolor[rgb]{0.00,0.27,0.87}{##1}}}
\expandafter\def\csname PYG@tok@err\endcsname{\def\PYG@bc##1{\setlength{\fboxsep}{0pt}\fcolorbox[rgb]{1.00,0.00,0.00}{1,1,1}{\strut ##1}}}
\expandafter\def\csname PYG@tok@kc\endcsname{\let\PYG@bf=\textbf\def\PYG@tc##1{\textcolor[rgb]{0.00,0.50,0.00}{##1}}}
\expandafter\def\csname PYG@tok@kd\endcsname{\let\PYG@bf=\textbf\def\PYG@tc##1{\textcolor[rgb]{0.00,0.50,0.00}{##1}}}
\expandafter\def\csname PYG@tok@kn\endcsname{\let\PYG@bf=\textbf\def\PYG@tc##1{\textcolor[rgb]{0.00,0.50,0.00}{##1}}}
\expandafter\def\csname PYG@tok@kr\endcsname{\let\PYG@bf=\textbf\def\PYG@tc##1{\textcolor[rgb]{0.00,0.50,0.00}{##1}}}
\expandafter\def\csname PYG@tok@bp\endcsname{\def\PYG@tc##1{\textcolor[rgb]{0.00,0.50,0.00}{##1}}}
\expandafter\def\csname PYG@tok@fm\endcsname{\def\PYG@tc##1{\textcolor[rgb]{0.00,0.00,1.00}{##1}}}
\expandafter\def\csname PYG@tok@vc\endcsname{\def\PYG@tc##1{\textcolor[rgb]{0.10,0.09,0.49}{##1}}}
\expandafter\def\csname PYG@tok@vg\endcsname{\def\PYG@tc##1{\textcolor[rgb]{0.10,0.09,0.49}{##1}}}
\expandafter\def\csname PYG@tok@vi\endcsname{\def\PYG@tc##1{\textcolor[rgb]{0.10,0.09,0.49}{##1}}}
\expandafter\def\csname PYG@tok@vm\endcsname{\def\PYG@tc##1{\textcolor[rgb]{0.10,0.09,0.49}{##1}}}
\expandafter\def\csname PYG@tok@sa\endcsname{\def\PYG@tc##1{\textcolor[rgb]{0.73,0.13,0.13}{##1}}}
\expandafter\def\csname PYG@tok@sb\endcsname{\def\PYG@tc##1{\textcolor[rgb]{0.73,0.13,0.13}{##1}}}
\expandafter\def\csname PYG@tok@sc\endcsname{\def\PYG@tc##1{\textcolor[rgb]{0.73,0.13,0.13}{##1}}}
\expandafter\def\csname PYG@tok@dl\endcsname{\def\PYG@tc##1{\textcolor[rgb]{0.73,0.13,0.13}{##1}}}
\expandafter\def\csname PYG@tok@s2\endcsname{\def\PYG@tc##1{\textcolor[rgb]{0.73,0.13,0.13}{##1}}}
\expandafter\def\csname PYG@tok@sh\endcsname{\def\PYG@tc##1{\textcolor[rgb]{0.73,0.13,0.13}{##1}}}
\expandafter\def\csname PYG@tok@s1\endcsname{\def\PYG@tc##1{\textcolor[rgb]{0.73,0.13,0.13}{##1}}}
\expandafter\def\csname PYG@tok@mb\endcsname{\def\PYG@tc##1{\textcolor[rgb]{0.40,0.40,0.40}{##1}}}
\expandafter\def\csname PYG@tok@mf\endcsname{\def\PYG@tc##1{\textcolor[rgb]{0.40,0.40,0.40}{##1}}}
\expandafter\def\csname PYG@tok@mh\endcsname{\def\PYG@tc##1{\textcolor[rgb]{0.40,0.40,0.40}{##1}}}
\expandafter\def\csname PYG@tok@mi\endcsname{\def\PYG@tc##1{\textcolor[rgb]{0.40,0.40,0.40}{##1}}}
\expandafter\def\csname PYG@tok@il\endcsname{\def\PYG@tc##1{\textcolor[rgb]{0.40,0.40,0.40}{##1}}}
\expandafter\def\csname PYG@tok@mo\endcsname{\def\PYG@tc##1{\textcolor[rgb]{0.40,0.40,0.40}{##1}}}
\expandafter\def\csname PYG@tok@ch\endcsname{\let\PYG@it=\textit\def\PYG@tc##1{\textcolor[rgb]{0.25,0.50,0.50}{##1}}}
\expandafter\def\csname PYG@tok@cm\endcsname{\let\PYG@it=\textit\def\PYG@tc##1{\textcolor[rgb]{0.25,0.50,0.50}{##1}}}
\expandafter\def\csname PYG@tok@cpf\endcsname{\let\PYG@it=\textit\def\PYG@tc##1{\textcolor[rgb]{0.25,0.50,0.50}{##1}}}
\expandafter\def\csname PYG@tok@c1\endcsname{\let\PYG@it=\textit\def\PYG@tc##1{\textcolor[rgb]{0.25,0.50,0.50}{##1}}}
\expandafter\def\csname PYG@tok@cs\endcsname{\let\PYG@it=\textit\def\PYG@tc##1{\textcolor[rgb]{0.25,0.50,0.50}{##1}}}

\def\PYGZbs{\char`\\}
\def\PYGZus{\char`\_}
\def\PYGZob{\char`\{}
\def\PYGZcb{\char`\}}

\def\PYGZam{\char`\&}
\def\PYGZlt{\char`\<}
\def\PYGZgt{\char`\>}
\def\PYGZsh{\char`\#}

\def\PYGZhy{\char`\-}
\def\PYGZsq{\char`\'}
\def\PYGZdq{\char`\"}
\def\PYGZti{\char`\~}
\makeatother

\makeatletter
\def\PYGdefault@reset{\let\PYGdefault@it=\relax \let\PYGdefault@bf=\relax%
    \let\PYGdefault@ul=\relax \let\PYGdefault@tc=\relax%
    \let\PYGdefault@bc=\relax \let\PYGdefault@ff=\relax}
\def\PYGdefault@tok#1{\csname PYGdefault@tok@#1\endcsname}
\def\PYGdefault@toks#1+{\ifx\relax#1\empty\else%
    \PYGdefault@tok{#1}\expandafter\PYGdefault@toks\fi}
\def\PYGdefault@do#1{\PYGdefault@bc{\PYGdefault@tc{\PYGdefault@ul{%
    \PYGdefault@it{\PYGdefault@bf{\PYGdefault@ff{#1}}}}}}}
\def\PYGdefault#1#2{\PYGdefault@reset\PYGdefault@toks#1+\relax+\PYGdefault@do{#2}}

\expandafter\def\csname PYGdefault@tok@w\endcsname{\def\PYGdefault@tc##1{\textcolor[rgb]{0.73,0.73,0.73}{##1}}}
\expandafter\def\csname PYGdefault@tok@c\endcsname{\let\PYGdefault@it=\textit\def\PYGdefault@tc##1{\textcolor[rgb]{0.25,0.50,0.50}{##1}}}
\expandafter\def\csname PYGdefault@tok@cp\endcsname{\def\PYGdefault@tc##1{\textcolor[rgb]{0.74,0.48,0.00}{##1}}}
\expandafter\def\csname PYGdefault@tok@k\endcsname{\let\PYGdefault@bf=\textbf\def\PYGdefault@tc##1{\textcolor[rgb]{0.00,0.50,0.00}{##1}}}
\expandafter\def\csname PYGdefault@tok@kp\endcsname{\def\PYGdefault@tc##1{\textcolor[rgb]{0.00,0.50,0.00}{##1}}}
\expandafter\def\csname PYGdefault@tok@kt\endcsname{\def\PYGdefault@tc##1{\textcolor[rgb]{0.69,0.00,0.25}{##1}}}
\expandafter\def\csname PYGdefault@tok@o\endcsname{\def\PYGdefault@tc##1{\textcolor[rgb]{0.40,0.40,0.40}{##1}}}
\expandafter\def\csname PYGdefault@tok@ow\endcsname{\let\PYGdefault@bf=\textbf\def\PYGdefault@tc##1{\textcolor[rgb]{0.67,0.13,1.00}{##1}}}
\expandafter\def\csname PYGdefault@tok@nb\endcsname{\def\PYGdefault@tc##1{\textcolor[rgb]{0.00,0.50,0.00}{##1}}}
\expandafter\def\csname PYGdefault@tok@nf\endcsname{\def\PYGdefault@tc##1{\textcolor[rgb]{0.00,0.00,1.00}{##1}}}
\expandafter\def\csname PYGdefault@tok@nc\endcsname{\let\PYGdefault@bf=\textbf\def\PYGdefault@tc##1{\textcolor[rgb]{0.00,0.00,1.00}{##1}}}
\expandafter\def\csname PYGdefault@tok@nn\endcsname{\let\PYGdefault@bf=\textbf\def\PYGdefault@tc##1{\textcolor[rgb]{0.00,0.00,1.00}{##1}}}
\expandafter\def\csname PYGdefault@tok@ne\endcsname{\let\PYGdefault@bf=\textbf\def\PYGdefault@tc##1{\textcolor[rgb]{0.82,0.25,0.23}{##1}}}
\expandafter\def\csname PYGdefault@tok@nv\endcsname{\def\PYGdefault@tc##1{\textcolor[rgb]{0.10,0.09,0.49}{##1}}}
\expandafter\def\csname PYGdefault@tok@no\endcsname{\def\PYGdefault@tc##1{\textcolor[rgb]{0.53,0.00,0.00}{##1}}}
\expandafter\def\csname PYGdefault@tok@nl\endcsname{\def\PYGdefault@tc##1{\textcolor[rgb]{0.63,0.63,0.00}{##1}}}
\expandafter\def\csname PYGdefault@tok@ni\endcsname{\let\PYGdefault@bf=\textbf\def\PYGdefault@tc##1{\textcolor[rgb]{0.60,0.60,0.60}{##1}}}
\expandafter\def\csname PYGdefault@tok@na\endcsname{\def\PYGdefault@tc##1{\textcolor[rgb]{0.49,0.56,0.16}{##1}}}
\expandafter\def\csname PYGdefault@tok@nt\endcsname{\let\PYGdefault@bf=\textbf\def\PYGdefault@tc##1{\textcolor[rgb]{0.00,0.50,0.00}{##1}}}
\expandafter\def\csname PYGdefault@tok@nd\endcsname{\def\PYGdefault@tc##1{\textcolor[rgb]{0.67,0.13,1.00}{##1}}}
\expandafter\def\csname PYGdefault@tok@s\endcsname{\def\PYGdefault@tc##1{\textcolor[rgb]{0.73,0.13,0.13}{##1}}}
\expandafter\def\csname PYGdefault@tok@sd\endcsname{\let\PYGdefault@it=\textit\def\PYGdefault@tc##1{\textcolor[rgb]{0.73,0.13,0.13}{##1}}}
\expandafter\def\csname PYGdefault@tok@si\endcsname{\let\PYGdefault@bf=\textbf\def\PYGdefault@tc##1{\textcolor[rgb]{0.73,0.40,0.53}{##1}}}
\expandafter\def\csname PYGdefault@tok@se\endcsname{\let\PYGdefault@bf=\textbf\def\PYGdefault@tc##1{\textcolor[rgb]{0.73,0.40,0.13}{##1}}}
\expandafter\def\csname PYGdefault@tok@sr\endcsname{\def\PYGdefault@tc##1{\textcolor[rgb]{0.73,0.40,0.53}{##1}}}
\expandafter\def\csname PYGdefault@tok@ss\endcsname{\def\PYGdefault@tc##1{\textcolor[rgb]{0.10,0.09,0.49}{##1}}}
\expandafter\def\csname PYGdefault@tok@sx\endcsname{\def\PYGdefault@tc##1{\textcolor[rgb]{0.00,0.50,0.00}{##1}}}
\expandafter\def\csname PYGdefault@tok@m\endcsname{\def\PYGdefault@tc##1{\textcolor[rgb]{0.40,0.40,0.40}{##1}}}
\expandafter\def\csname PYGdefault@tok@gh\endcsname{\let\PYGdefault@bf=\textbf\def\PYGdefault@tc##1{\textcolor[rgb]{0.00,0.00,0.50}{##1}}}
\expandafter\def\csname PYGdefault@tok@gu\endcsname{\let\PYGdefault@bf=\textbf\def\PYGdefault@tc##1{\textcolor[rgb]{0.50,0.00,0.50}{##1}}}
\expandafter\def\csname PYGdefault@tok@gd\endcsname{\def\PYGdefault@tc##1{\textcolor[rgb]{0.63,0.00,0.00}{##1}}}
\expandafter\def\csname PYGdefault@tok@gi\endcsname{\def\PYGdefault@tc##1{\textcolor[rgb]{0.00,0.63,0.00}{##1}}}
\expandafter\def\csname PYGdefault@tok@gr\endcsname{\def\PYGdefault@tc##1{\textcolor[rgb]{1.00,0.00,0.00}{##1}}}
\expandafter\def\csname PYGdefault@tok@ge\endcsname{\let\PYGdefault@it=\textit}
\expandafter\def\csname PYGdefault@tok@gs\endcsname{\let\PYGdefault@bf=\textbf}
\expandafter\def\csname PYGdefault@tok@gp\endcsname{\let\PYGdefault@bf=\textbf\def\PYGdefault@tc##1{\textcolor[rgb]{0.00,0.00,0.50}{##1}}}
\expandafter\def\csname PYGdefault@tok@go\endcsname{\def\PYGdefault@tc##1{\textcolor[rgb]{0.53,0.53,0.53}{##1}}}
\expandafter\def\csname PYGdefault@tok@gt\endcsname{\def\PYGdefault@tc##1{\textcolor[rgb]{0.00,0.27,0.87}{##1}}}
\expandafter\def\csname PYGdefault@tok@err\endcsname{\def\PYGdefault@bc##1{\setlength{\fboxsep}{0pt}\fcolorbox[rgb]{1.00,0.00,0.00}{1,1,1}{\strut ##1}}}
\expandafter\def\csname PYGdefault@tok@kc\endcsname{\let\PYGdefault@bf=\textbf\def\PYGdefault@tc##1{\textcolor[rgb]{0.00,0.50,0.00}{##1}}}
\expandafter\def\csname PYGdefault@tok@kd\endcsname{\let\PYGdefault@bf=\textbf\def\PYGdefault@tc##1{\textcolor[rgb]{0.00,0.50,0.00}{##1}}}
\expandafter\def\csname PYGdefault@tok@kn\endcsname{\let\PYGdefault@bf=\textbf\def\PYGdefault@tc##1{\textcolor[rgb]{0.00,0.50,0.00}{##1}}}
\expandafter\def\csname PYGdefault@tok@kr\endcsname{\let\PYGdefault@bf=\textbf\def\PYGdefault@tc##1{\textcolor[rgb]{0.00,0.50,0.00}{##1}}}
\expandafter\def\csname PYGdefault@tok@bp\endcsname{\def\PYGdefault@tc##1{\textcolor[rgb]{0.00,0.50,0.00}{##1}}}
\expandafter\def\csname PYGdefault@tok@fm\endcsname{\def\PYGdefault@tc##1{\textcolor[rgb]{0.00,0.00,1.00}{##1}}}
\expandafter\def\csname PYGdefault@tok@vc\endcsname{\def\PYGdefault@tc##1{\textcolor[rgb]{0.10,0.09,0.49}{##1}}}
\expandafter\def\csname PYGdefault@tok@vg\endcsname{\def\PYGdefault@tc##1{\textcolor[rgb]{0.10,0.09,0.49}{##1}}}
\expandafter\def\csname PYGdefault@tok@vi\endcsname{\def\PYGdefault@tc##1{\textcolor[rgb]{0.10,0.09,0.49}{##1}}}
\expandafter\def\csname PYGdefault@tok@vm\endcsname{\def\PYGdefault@tc##1{\textcolor[rgb]{0.10,0.09,0.49}{##1}}}
\expandafter\def\csname PYGdefault@tok@sa\endcsname{\def\PYGdefault@tc##1{\textcolor[rgb]{0.73,0.13,0.13}{##1}}}
\expandafter\def\csname PYGdefault@tok@sb\endcsname{\def\PYGdefault@tc##1{\textcolor[rgb]{0.73,0.13,0.13}{##1}}}
\expandafter\def\csname PYGdefault@tok@sc\endcsname{\def\PYGdefault@tc##1{\textcolor[rgb]{0.73,0.13,0.13}{##1}}}
\expandafter\def\csname PYGdefault@tok@dl\endcsname{\def\PYGdefault@tc##1{\textcolor[rgb]{0.73,0.13,0.13}{##1}}}
\expandafter\def\csname PYGdefault@tok@s2\endcsname{\def\PYGdefault@tc##1{\textcolor[rgb]{0.73,0.13,0.13}{##1}}}
\expandafter\def\csname PYGdefault@tok@sh\endcsname{\def\PYGdefault@tc##1{\textcolor[rgb]{0.73,0.13,0.13}{##1}}}
\expandafter\def\csname PYGdefault@tok@s1\endcsname{\def\PYGdefault@tc##1{\textcolor[rgb]{0.73,0.13,0.13}{##1}}}
\expandafter\def\csname PYGdefault@tok@mb\endcsname{\def\PYGdefault@tc##1{\textcolor[rgb]{0.40,0.40,0.40}{##1}}}
\expandafter\def\csname PYGdefault@tok@mf\endcsname{\def\PYGdefault@tc##1{\textcolor[rgb]{0.40,0.40,0.40}{##1}}}
\expandafter\def\csname PYGdefault@tok@mh\endcsname{\def\PYGdefault@tc##1{\textcolor[rgb]{0.40,0.40,0.40}{##1}}}
\expandafter\def\csname PYGdefault@tok@mi\endcsname{\def\PYGdefault@tc##1{\textcolor[rgb]{0.40,0.40,0.40}{##1}}}
\expandafter\def\csname PYGdefault@tok@il\endcsname{\def\PYGdefault@tc##1{\textcolor[rgb]{0.40,0.40,0.40}{##1}}}
\expandafter\def\csname PYGdefault@tok@mo\endcsname{\def\PYGdefault@tc##1{\textcolor[rgb]{0.40,0.40,0.40}{##1}}}
\expandafter\def\csname PYGdefault@tok@ch\endcsname{\let\PYGdefault@it=\textit\def\PYGdefault@tc##1{\textcolor[rgb]{0.25,0.50,0.50}{##1}}}
\expandafter\def\csname PYGdefault@tok@cm\endcsname{\let\PYGdefault@it=\textit\def\PYGdefault@tc##1{\textcolor[rgb]{0.25,0.50,0.50}{##1}}}
\expandafter\def\csname PYGdefault@tok@cpf\endcsname{\let\PYGdefault@it=\textit\def\PYGdefault@tc##1{\textcolor[rgb]{0.25,0.50,0.50}{##1}}}
\expandafter\def\csname PYGdefault@tok@c1\endcsname{\let\PYGdefault@it=\textit\def\PYGdefault@tc##1{\textcolor[rgb]{0.25,0.50,0.50}{##1}}}
\expandafter\def\csname PYGdefault@tok@cs\endcsname{\let\PYGdefault@it=\textit\def\PYGdefault@tc##1{\textcolor[rgb]{0.25,0.50,0.50}{##1}}}

\makeatother

\begin{document}

\title{XACC: A System-Level Software Infrastructure for Heterogeneous Quantum-Classical Computing}

\thanks{This manuscript has been authored by UT-Battelle, LLC under Contract No. DE-AC05-00OR22725 with the U.S. Department of Energy. The United States Government retains and the publisher, by accepting the article for publication, acknowledges that the United States Government retains a non-exclusive, paid-up, irrevocable, world-wide license to publish or reproduce the published form of this manuscript, or allow others to do so, for United States Government purposes. The Department of Energy will provide public access to these results of federally sponsored research in accordance with the DOE Public Access Plan. (http://energy.gov/downloads/doe-public-access-plan).}

\begin{abstract}
Quantum programming techniques and software have advanced significantly over the past five years, with a majority focusing on high-level language frameworks targeting remote REST library APIs. As quantum computing architectures advance and become more widely available, lower-level, system software infrastructures will be needed to enable tighter, co-processor programming and access models. Here we present XACC, a system-level software infrastructure for quantum-classical computing that promotes a service-oriented architecture to expose interfaces for core quantum programming, compilation, and execution tasks. We detail XACC's interfaces, their interactions, and its implementation as a hardware-agnostic framework for both near-term and future quantum-classical architectures. We provide concrete examples demonstrating the utility of this framework with paradigmatic tasks. Our approach lays the foundation for the development of compilers, associated runtimes, and low-level system tools tightly integrating quantum and classical workflows.
\end{abstract}

\author{Alexander J.\ McCaskey}
\email{mccaskeyaj@ornl.gov}
\affiliation{Quantum Computing Institute,\ Oak\ Ridge\ National\ Laboratory,\
  Oak\ Ridge,\ TN,\ 37831,\ USA}
\affiliation{Computer Science and Mathematics Division,\ Oak\ Ridge\ National\ Laboratory,\ Oak\ Ridge,\ TN,\ 37831,\ USA}

\author{Dmitry I. Lyakh}
\affiliation{Quantum Computing Institute,\ Oak\ Ridge\ National\ Laboratory,\
  Oak\ Ridge,\ TN,\ 37831,\ USA}
\affiliation{National Center for Computational Sciences,\ Oak\ Ridge\ National\ Laboratory,\ Oak\ Ridge,\ TN,\ 37831,\ USA}

\author{Eugene F. Dumitrescu}
\affiliation{Quantum Computing Institute,\ Oak\ Ridge\ National\ Laboratory,\
  Oak\ Ridge,\ TN,\ 37831,\ USA}
\affiliation{Computational Sciences and Engineering Division,\ Oak\ Ridge\ National\ Laboratory,\ Oak\ Ridge,\ TN,\ 37831,\ USA}

\author{Sarah S.\ Powers}
\affiliation{Quantum Computing Institute,\ Oak\ Ridge\ National\ Laboratory,\
  Oak\ Ridge,\ TN,\ 37831,\ USA}
\affiliation{Computer Science and Mathematics Division,\ Oak\ Ridge\ National\ Laboratory,\ Oak\ Ridge,\ TN,\ 37831,\ USA}

\author{Travis S. Humble}
\affiliation{Quantum Computing Institute,\ Oak\ Ridge\ National\ Laboratory,\
  Oak\ Ridge,\ TN,\ 37831,\ USA}
\affiliation{Computational Sciences and Engineering Division,\ Oak\ Ridge\ National\ Laboratory,\ Oak\ Ridge,\ TN,\ 37831,\ USA}

\maketitle

%%%%%%%%%%%%%%%%%%%%%%%%%
\section{Introduction}
The near-term availability of noisy quantum processing units (QPUs) has enabled a number of proof-of-principle demonstrations of quantum co-processing for existing domain computational science \cite{Dumitrescu2018, ornl-qchem, hamilton,Klco2018,PhysRevA.99.032306}. These demonstrations take advantage of sophisticated software frameworks enabling programming, compilation, and execution of quantum algorithms on remotely hosted QPUs. The majority of these efforts have thus far focused on the implementation of high-level, interpreted languages that enable quantum circuit composition, transpilation, and vendor-specific back-end execution \cite{Qiskit,cirq,quil,openfermionarxiv}. These programming frameworks are well suited for near-term experimentation on remotely hosted quantum resources, as they enable quick prototyping and rely solely on existing remote communication libraries.
\par
As QPUs scale, system designs are expected to more tightly integrate CPU and QPU interactions \cite{britt2017high,McCaskeyICRC2018}. A quantum-accelerated computing model should adopt best practices and mirror the design of modern heterogeneous high-performance computing, which offers the advantage of improved performance through more refined control over execution scheduling and memory management. Along with tighter integration, the use of interpreted languages for controlling remote access execution must be replaced by lower-level, system software infrastructures, tools, and compilers for composing complex quantum-classical workflows. System-level software is necessary for robust mechanisms allowing quantum device driver execution as part of a tightly integrated co-processor programming model. Single-source quantum-classical compilers and tools, as well as methods for benchmarking, profiling, and debugging, will also be important for enabling domain computational science on these next-generation heterogeneous systems. Initial models for quantum system infrastructure can be realized using system-level languages such as C and C\texttt{++} that have served as the foundation for high-performance application development on conventional hardware. Native languages may also expose bindings for higher-level languages, like Python, to enable experimentation and productivity.
\par
To achieve these goals, we present the XACC system-level software framework. XACC provides an open-source, low-level programming framework for the development of quantum-accelerated programs. The framework implements a hardware-agnostic approach to device integration that allows programmers to target multiple concrete hardware back-ends. XACC is designed around a familiar co-processor programming model that leverages an efficient, extensible, and modular service-oriented architecture. This is accomplished using a plug-and-play capability for a holistic quantum programming, compilation, and execution workflow.
\par
Our presentation of the XACC programming framework is organized as follows: in Sec.~\ref{sec:goals} we summarize the salient features of XACC. In Sec.~\ref{sec:arch} we detail the underlying service-oriented architecture and implementations. We describe the language interfaces for XACC in Sec.~\ref{sec:inter}. We provide example demonstrations in Sec.~\ref{sec:demo} and offer conclusions in Sec.~\ref{sec:disc}.
%%%
\section{Features of XACC}
\label{sec:goals}
The purpose of the XACC framework is to provide a holistic, low-level software infrastructure that enables cross-platform programming, compilation, and execution of hybrid quantum-classical scientific applications. The framework adheres to the following design goals:
\begin{itemize}[leftmargin=*]

\item[$1)$] \emph{Provide a quantum co-processor programming model} The XACC infrastructure manages QPUs as accelerators or co-processors to conventional (classical) computing systems. This design assumes that traditional scientific use cases will leverage QPUs by off-loading select classically intractable kernels to the available quantum resource and later process the results to inform remaining classical computations. We design the XACC framework to provide an extensible quantum accelerator back-end that defines common operations and communication protocols with a general, abstract quantum computer.

\item[$2)$] \emph{Provide low-level system software interfaces} XACC exposes interfaces for the low-level system control of the CPU and QPU. These interfaces support a diversity of infrastructure tools, approaches, compilers, and languages for quantum-accelerated computing. The XACC framework is implemented as a C\texttt{++}
%(C\texttt{++}11)
infrastructure in order to integrate existing quantum programming methods. Starting with a low-level, ubiquitous language like C\texttt{++} enables extension to higher-level languages, such as Python, Julia, etc. through well-supported language binding libraries. C\texttt{++} is also well-known for its portability, performance, and efficiency for conventional computing systems, and it is highly suitable for future hybrid heterogeneous systems with tighter quantum-classical integration. We therefore use C\texttt{++} to lay the ground work for tighter integration models that move past current remote-process-invocation protocols to in-process, in-memory device driver access.

\item[$3)$] \emph{Support cross-platform operation} The XACC framework is designed to be agnostic with respect to the QPU hardware. Users are able to compose and compile applications for multiple quantum computing platforms using a single, common language. This feature supports portability as well as comparative analyses of commonly defined benchmarks. Since source cross-platform operation will prove critical in determining which hardware is best suited for distinct applications.

\item[$4)$] \emph{Support modular and extensible functionality} The XACC framework supports a service-oriented, or plugin, architecture that enables modular and extensible functionality. Such a service-oriented software architecture is useful in integrating hardware back-ends, high-level programming approaches, intermediate-level quantum compilation, and error mitigation strategies. A plugin architecture directly enables researchers and programmers to efficiently swap out key aspects of the overall programming, compilation, and execution workflow with problem-specific implementations.

%% This is implementation details
%To enable this design, we have focused on enabling a low-level, service-oriented plug-and-play architecture that borrows from the Java modular application programming world. We leverage the Open Service Gateway Initiative (OSGi) and corresponding native C\texttt{++} implementation, CppMicroServices, to decompose the quantum programming, compilation, and execution workflow into modular \emph{bundles} \fix{that expose framework interface implementations that contribute to this workflow -- 'that' used multiple times in a row makes things not read as smoothly}. These bundles are deployed as standard shared libraries and contributed to the framework at runtime.

\end{itemize}

These features distinguish XACC from other programming frameworks. First, XACC is the only quantum-classical programming framework that provides a system-level software infrastructure permitting closer CPU-QPU integration.
% Other programming frameworks focus on high-level Pythonic approaches to enabling remotely hosted CPU-QPU access models, or extend other high-level languages that are not common across the high-performance computing research space.
As such, we expect XACC to directly influence future QPU integration efforts for existing high-performance scientific computing codes written in Fortran, C, and C++.
% This will directly affect expected future QPU integration efforts for existing high-performance scientific computing codes, the majority of which are written in Fortran, C, and C++.
As it is not a vendor-supplied framework, XACC further differentiates itself by promoting device interoperability as a core feature and design goal.
By keeping the core XACC design abstract and extensible, it is also able to target different quantum computing models: gate-based and annealing.

\section{Framework Architecture}
\label{sec:arch}
\begin{figure}[!ht]
\centering
\includegraphics[width=\columnwidth]{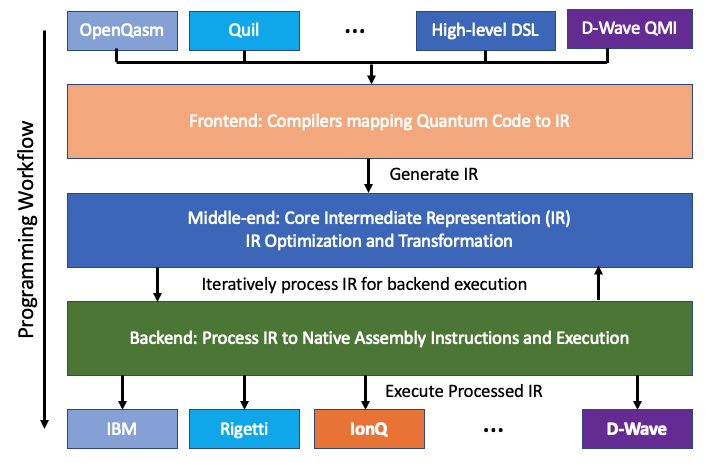}
% \missingfigure[figwidth=\columnwidth]{layers}
\caption{XACC is composed of three high-level layers - the front-end, middle-end, and back-end. The overall workflow starts with quantum kernel programming at the frontend, followed by IR generation and processing, and ends with back-end execution. Each of these layers exposes a variety of critical extension points.
}
\label{fig:layered}
\end{figure}

The architecture of the XACC programming framework exposes a series of interfaces for hardware-agnostic quantum programming, compilation, and execution.
As shown in Fig.~\ref{fig:layered}, the XACC framework uses a layered software architecture based upon the common compiler decomposition of a front-end, middle-end, and back-end layers. Each layer exposes different extension points, or interfaces, for implementing a variety of specific use cases. XACC currently includes interfaces for quantum language compilers, instructions, hardware devices (also referred to as accelerators), compiler optimizations (alternatively, passes), observable measurements, and algorithms.
%%
%\subsection{Overview of XACC Layers}
\par
The front-end maps quantum code expressions, i.e. source code, into an XACC intermediate representation (IR) suitable for transformations and analysis by the middle layer. More specifically, the front-end layer maps source strings expressing a \emph{quantum kernel} to an IR instance. For example, the XACC front-end may be extended to parse an input source code written in the IBM OpenQASM \cite{openqasm} dialect and enable translation into the XACC IR. Similar functionality is implemented for the Rigetti Quil \cite{quil} dialect and higher-level languages. As a framework, XACC provides an extensible quantum code transpilation or compilation interface that may be tailored to parse many different languages into a common IR for subsequent manipulation.
%in an extensible and modular fashion (see Sec. \ref{sec:compilers}).
\par
The middle layer exposes an IR object model while maintaining an application programming interface (API) for quantum programs that is agnostic to the underlying hardware. The structure of the IR provides an integration mechanism generalizing disparate programming approaches with multiple quantum devices. A critical middle layer concept is the IR transformation, which exposes a framework extension point for performing standard quantum compilation routines, including circuit synthesis, as well as the addition of error mitigation and correction techniques into the logical design.
\par
The back-end layer exposes an abstract quantum computer interface that accepts instances of the IR and executes them on a targeted hardware device (see Sec. \ref{sec:acc}). The back-end layer provides further extension points mapping the IR to hardware-specific instruction sets and the low-level controls used to execute kernels. The former is performed by taking advantage of the middle-end's IR transformation infrastructure. By using a common interface, quantum program execution via the back-end layer is easily extensible to new hardware. % Extensibility of the back-end layer uses a common interface to execute programs directly on quantum hardware.
\par
% This layered architecture is that it promotes an overall separation of concerns that enables management of the complexity of applying quantum programming methods across the high and low-level language dialects native to specific quantum computing devices.
With the layered architecture, XACC efficiently maps high-level programming languages to low-level hardware instruction sets. Where a direct mapping of $N$ languages into $M$ quantum devices would require $NM$ separate language-to-hardware mappings, XACC reduces the amount of work necessary to $N+M$ separate mapping implementations. The IR is the central construct connecting the front-end and back-end layers through a set of transformations.
% This is a consequence of providing common extension points between frontend and back-end layers.
% Central to this concept is the definition of a common IR for communication between transformations.

\subsection{Accelerator Buffer}
\begin{figure}[!t]
\centering
\includegraphics[width=\columnwidth]{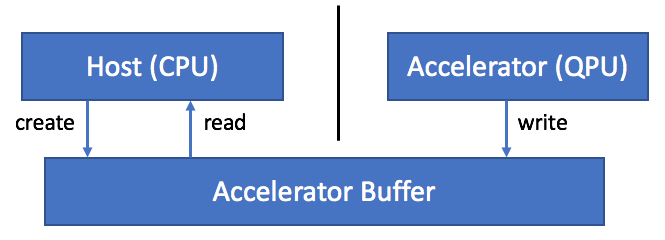}
% \missingfigure[figwidth=\columnwidth]{accbuf}
\caption{The Accelerator Buffer serves as a memory object for addressing a quantum register and provides a mechanism for posting and retrieving execution results and associated metadata. XACC manages the interactions between the host system and the QPU to track the state of the Accelerator Buffer for both  remote and local execution models.}
\label{fig:accb}
\end{figure}
XACC models general quantum programs as operations on an allocated register of qubits. Specifically, XACC puts forward an \emph{accelerator buffer} abstraction that models the underlying quantum register, or buffer, on which quantum operations are performed. As a memory buffer, it is allocated at run-time by the programmer to define the size of the register and to store results corresponding to measurement of the individual register elements. Subsequent execution of the resulting program will generate measurement results and metadata that are assigned to that buffer instance. XACC manages the reference to accelerator buffers throughout program execution and gathers the measurement results for assignment to the buffer. Since programmers allocate buffers, and buffers are operated on by some execution back-end, the buffer concept spans the front-end, middle-end, and back-end layers. Associated metadata tracks which qubits are operated upon as well as the aggregation of quantum execution measurement results.
\par
\begin{figure}[!t]
\centering
\includegraphics[width=\columnwidth]{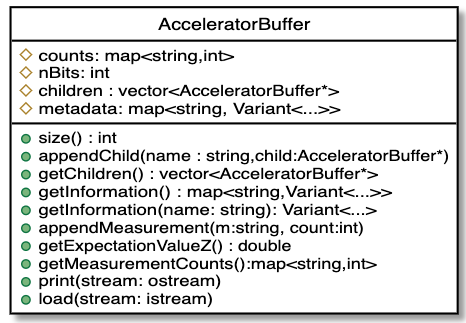}
% \missingfigure[figwidth=\columnwidth]{accbuf}
\caption{\texttt{AcceleratorBuffer} provides a data structure for addressing a quantum register and storing measurement results. The class provides a mechanism for posting and retrieving execution results and associated metadata.}
\label{fig:accbclass}
\end{figure}

To model the buffer abstraction, XACC defines an \texttt{AcceleratorBuffer} class that programmers must instantiate, containing at least the number of register elements required by the program. This buffer is passed by reference to the XACC accelerator execution back-end, which provides the execution infrastructure (see Sec.~\ref{sec:acc}). Methods and members of the \texttt{AcceleratorBuffer} base class are shown in Fig.~\ref{fig:accbclass}.
\par
An \texttt{AcceleratorBuffer} tracks the mapping of measured bit strings to the number of times they were observed (the \texttt{counts} map in Fig. \ref{fig:accbclass}), as well as an associated heterogeneous map (described in Sec.~\ref{sec:utils}) of metadata. The buffer provides measurement results to the programmer by reference as shown in Fig.~\ref{fig:accb}. The metadata maps string keys to a variant value type, which includes typical C\texttt{++} types like integer, floating point, string, and vector data structures. The metadata member is intended to store execution-pertinent information that may be different across available quantum computing back-ends. For example, it may be populated by the back-end infrastructure with information about device properties such as noise, as well as variational parameters corresponding to the execution and expectation values.
\par
The \texttt{AcceleratorBuffer} class tracks a list of children buffers and thereby enables a tree of execution data. This feature is well suited for variational methods, in which an initial buffer is appended as the subsequent executions are updated based on circuit refinement.
%single \texttt{AcceleratorBuffer} for the variational problem at hand, pass that along for iterative execution of the quantum back-end, and
New children buffers are appended to the global root buffer with each iteration, such that each child owns metadata and measurement results describing the iteration.
\par
The code snippet in Fig. \ref{fig:accb_code} shows a buffer being allocated, passed to the back-end execution infrastructure, and its results accessed.
\begin{figure}[!b]
\centering
\begin{tcolorbox}[colback=white]
\begin{Verbatim}[commandchars=\\\{\}]
\PYG{c+c1}{// User allocates buffer}
\PYG{c+c1}{// keeps reference to it throughout execution}
\PYG{k}{auto} \PYG{n}{buffer} \PYG{o}{=} \PYG{n}{xacc}\PYG{o}{::}\PYG{n}{qalloc}\PYG{p}{(}\PYG{l+m+mi}{3}\PYG{p}{);}
\PYG{c+c1}{// Execute some circuit / composite instruction}
\PYG{c+c1}{// this writes result data to the buffer}
\PYG{n}{accelerator}\PYG{o}{\PYGZhy{}\PYGZgt{}}\PYG{n}{execute}\PYG{p}{(}\PYG{n}{buffer}\PYG{p}{,} \PYG{n}{circuit}\PYG{p}{);}
\PYG{c+c1}{// User still has that buffer, get results}
\PYG{k}{auto} \PYG{n}{results} \PYG{o}{=} \PYG{n}{buffer}\PYG{o}{\PYGZhy{}\PYGZgt{}}\PYG{n}{getMeasurementCounts}\PYG{p}{();}

\PYG{c+c1}{// Can add/get metadata, measurements,}
\PYG{c+c1}{// and expectation values}
\PYG{n}{std}\PYG{o}{::}\PYG{n}{vector}\PYG{o}{\PYGZlt{}}\PYG{k+kt}{double}\PYG{o}{\PYGZgt{}} \PYG{n}{fidelities} \PYG{o}{=}
 \PYG{n}{buffer}\PYG{o}{\PYGZhy{}\PYGZgt{}}\PYG{n}{getInformation}\PYG{p}{(}\PYG{l+s}{\PYGZdq{}1q\PYGZhy{}gate\PYGZhy{}fidelities\PYGZdq{}}\PYG{p}{)}
        \PYG{p}{.}\PYG{n}{as}\PYG{o}{\PYGZlt{}}\PYG{n}{std}\PYG{o}{::}\PYG{n}{vector}\PYG{o}{\PYGZlt{}}\PYG{k+kt}{double}\PYG{o}{\PYGZgt{}\PYGZgt{}}\PYG{p}{();}
\PYG{k}{auto} \PYG{n}{counts} \PYG{o}{=} \PYG{n}{buffer}\PYG{o}{\PYGZhy{}\PYGZgt{}}\PYG{n}{getMeasurementCounts}\PYG{p}{();}
\PYG{k}{auto} \PYG{n}{expVal} \PYG{o}{=} \PYG{n}{buffer}\PYG{o}{\PYGZhy{}\PYGZgt{}}\PYG{n}{getExpectationValueZ}\PYG{p}{();}
\end{Verbatim}
\end{tcolorbox}
\caption{Example of allocating a three qubit \texttt{AcceleratorBuffer} and executing a compiled circuit which persists results and metadata to the buffer. Note the \texttt{qalloc()} utility function for allocating quantum memory (AcceleratorBuffer).}
\label{fig:accb_code}
\end{figure}

\subsection{Kernels and Compilers}
\label{sec:compilers}
The XACC programming model stipulates that quantum code be described as standard C-like functions, called \emph{quantum kernels}. Kernels must adhere to the the following requirements:
\begin{itemize}
    \item \emph{Annotation} - XACC quantum kernels must be annotated with the \texttt{\_\_qpu\_\_} function attribute to enable static, ahead-of-time compilation.
    \item \emph{Unique Name} - Like standard C/C++ functions, kernels must have a unique function name.
    \item \emph{AcceleratorBuffer} - Kernels must take an \texttt{AcceleratorBuffer} as their first argument.
    \item \emph{Parameters} - Kernels can take any number of arguments after the \texttt{AcceleratorBuffer}.
    \item \emph{Language} - The kernel function body must be written in a valid language, i.e. there must exists an XACC \texttt{Compiler} implementation for that language.
\end{itemize}
\begin{figure}[!b]
\centering
\begin{tcolorbox}[colback=white]
\begin{Verbatim}[commandchars=\\\{\}]
\PYG{c+c1}{// Annotated}
\PYG{c+c1}{// AcceleratorBuffer first}
\PYG{c+c1}{// parameters after}
\PYG{n}{\PYGZus{}\PYGZus{}qpu\PYGZus{}\PYGZus{}} \PYG{n+nf}{kernel}\PYG{p}{(}\PYG{n}{AcceleratorBuffer} \PYG{n}{q}\PYG{p}{,} \PYG{k+kt}{double} \PYG{n}{x}\PYG{p}{)} \PYG{p}{\PYGZob{}}
   \PYG{c+c1}{// Function body written in}
   \PYG{c+c1}{// some \PYGZsq{}known\PYGZsq{} language, here xasm}
   \PYG{n}{X}\PYG{p}{(}\PYG{n}{q}\PYG{p}{[}\PYG{l+m+mi}{0}\PYG{p}{]);}
   \PYG{n}{Ry}\PYG{p}{(}\PYG{n}{q}\PYG{p}{[}\PYG{l+m+mi}{1}\PYG{p}{],} \PYG{n}{x}\PYG{p}{);}
   \PYG{n}{CX}\PYG{p}{(}\PYG{n}{q}\PYG{p}{[}\PYG{l+m+mi}{1}\PYG{p}{],} \PYG{n}{q}\PYG{p}{[}\PYG{l+m+mi}{0}\PYG{p}{]);}
   \PYG{n}{Measure}\PYG{p}{(}\PYG{n}{q}\PYG{p}{[}\PYG{l+m+mi}{0}\PYG{p}{]);}
\PYG{p}{\PYGZcb{}}
\end{Verbatim}

\end{tcolorbox}
\caption{Prototypical XACC Quantum Kernel with function body written in the default XACC assembly language (XASM, see Sec.~\ref{sec:inter})}
\label{fig:qcor_code}
\end{figure}

The final requirement introduces another abstraction that XACC exposes - the \texttt{Compiler} interface. \texttt{Compilers} take kernel source strings and map them to the XACC intermediate representation. This compile step takes the source string itself and the back-end accelerator that the programmer is compiling to, thereby enabling connectivity information, noise models, etc. at compile time. \texttt{Compilers} can be implemented for any number of languages with varying levels of language abstraction (low-level assembly, high-level domain specific languages, etc.), but must be able to parse the argument structure and function body to a valid instance of the IR. Since \texttt{Compilers} create IR from language source strings, they are also in position to map IR back to the their specific source language. XACC \texttt{Compilers} expose IR translation capabilities that enable this, thereby providing quantum source-to-source translation (for instance, mapping Quil to OpenQasm).

Using the base XACC API, programmers construct quantum kernel source code as standard strings or string literals, then get reference to the correct \texttt{Compiler} implementation for the source string, and compile it to an instance of the XACC IR. This is demonstrated in the code snippet in Fig. \ref{fig:compile}.

\subsection{Intermediate Representation}
\label{sec:ir}
\begin{figure*}[!t]
\centering
\includegraphics[width=\textwidth]{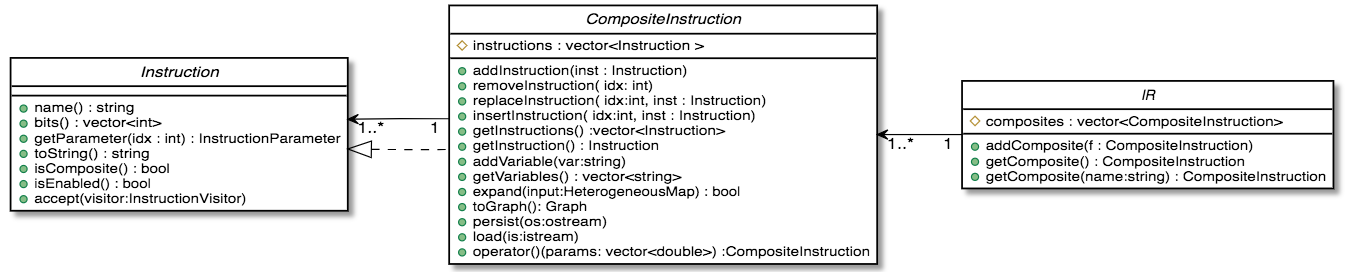}
% \missingfigure[figwidth=\columnwidth]{accbuf}
\caption{The XACC IR architecture, composed of \texttt{Instruction}, \texttt{CompositeInstruction}, and \texttt{IR} interfaces, enables a polymorphic object-model for the description of quantum programs that spans available quantum back-ends.}
\label{fig:ir_arch}
\end{figure*}

The XACC intermediate representation provides the architectural glue that connects the front-end programmer to back-end native assembly execution. The IR is a polymorphic data model residing at a slightly higher-level of abstraction than typical quantum assembly (QASM) and provides a manipulable, in-memory representation of quantum programs that is amenable to transformation and optimization. Moreover, a standard IR efficiently integrates multiple languages with various hardware instruction sets. The XACC IR architecture is shown in Fig. \ref{fig:ir_arch} and consists of three primary interfaces: \texttt{Instruction}, \texttt{CompositeInstruction}, and \texttt{IR}.

The IR begins with the definition of a general QASM-level instruction. The XACC \texttt{Instruction} interface describes general assembly instructions that have a unique name, operate on a set of qubits, are parameterized by concrete or variable parameters, and can be enabled/disabled. This interface is implemented for a set of instructions that are common in the digital gate or annealing models of quantum computation. Parameterized instructions are enabled via a variant type called \texttt{InstructionParameter}, which can be any of \texttt{int}, \texttt{double}, or \texttt{string} types. This allows instructions to depend on runtime parameters, providing a compile-once-and-update approach for iterative algorithms.

To compose instructions, XACC defines a \texttt{CompositeInstruction} interface which inherits from \texttt{Instruction}, but also contains a list of children instructions, modeling the familiar composite, or tree design pattern \cite{gof}. \texttt{CompositeInstructions} are therefore \emph{n-ary} trees where nodes are \texttt{CompositeInstructions} and leaves are concrete \texttt{Instructions}. \texttt{CompositeInstructions} expose methods for adding, removing, and replacing children instructions, as well as keeping track of any variables that children instructions may depend on. These variables are represented as string \texttt{InstructionParameter} instances. For example \texttt{Ry(t0)} on qubit $0$ depends on the \texttt{InstructionParameter} \texttt{t0}, thereby requiring any \texttt{CompositeInstruction} containing it to know about \texttt{t0}.

\begin{figure}[!t]
\centering
\begin{tcolorbox}[colback=white]
\begin{Verbatim}[commandchars=\\\{\}]
\PYG{k}{auto} \PYG{n}{qpu} \PYG{o}{=} \PYG{n}{xacc}\PYG{o}{::}\PYG{n}{getAccelerator}\PYG{p}{(}\PYG{l+s}{\PYGZdq{}ibm:back\PYGZhy{}end\PYGZdq{}}\PYG{p}{);}
\PYG{k}{auto} \PYG{n}{quil} \PYG{o}{=} \PYG{n}{xacc}\PYG{o}{::}\PYG{n}{getCompiler}\PYG{p}{(}\PYG{l+s}{\PYGZdq{}quil\PYGZdq{}}\PYG{p}{);}
\PYG{k}{auto} \PYG{n}{ir} \PYG{o}{=} \PYG{n}{quil}\PYG{o}{\PYGZhy{}\PYGZgt{}}\PYG{n}{compile}\PYG{p}{(}\PYG{l+s+sa}{R}\PYG{l+s}{\PYGZdq{}}\PYG{l+s+dl}{(}
\PYG{l+s}{\PYGZus{}\PYGZus{}qpu\PYGZus{}\PYGZus{} ansatz(AcceleratorBuffer q, double x)}
\PYG{l+s}{\PYGZob{}}
\PYG{l+s}{   X 0}
\PYG{l+s}{   Ry(x) 1}
\PYG{l+s}{   CX 1 0}
\PYG{l+s}{\PYGZcb{}}
\PYG{l+s}{\PYGZus{}\PYGZus{}qpu\PYGZus{}\PYGZus{} X0X1(AcceleratorBuffer q, double x)}
\PYG{l+s}{\PYGZob{}}
\PYG{l+s}{   ansatz(q, x);}
\PYG{l+s}{   H 0}
\PYG{l+s}{   H 1}
\PYG{l+s}{   MEASURE 0 [0]}
\PYG{l+s}{   MEASURE 1 [1]}
\PYG{l+s}{\PYGZcb{}}
\PYG{l+s+dl}{)}\PYG{l+s}{\PYGZdq{}}\PYG{p}{,} \PYG{n}{qpu}\PYG{p}{);}
\PYG{k}{auto} \PYG{n}{x0x1} \PYG{o}{=} \PYG{n}{ir}\PYG{o}{\PYGZhy{}\PYGZgt{}}\PYG{n}{getComposite}\PYG{p}{(}\PYG{l+s}{\PYGZdq{}X0X1\PYGZdq{}}\PYG{p}{);}

\PYG{c+c1}{// Translate quil kernel to openqasm...}
\PYG{k}{auto} \PYG{n}{openqasm} \PYG{o}{=} \PYG{n}{xacc}\PYG{o}{::}\PYG{n}{getCompiler}\PYG{p}{(}\PYG{l+s}{\PYGZdq{}openqasm\PYGZdq{}}\PYG{p}{);}
\PYG{k}{auto} \PYG{n}{oqasmSrc} \PYG{o}{=} \PYG{n}{openqasm}\PYG{o}{\PYGZhy{}\PYGZgt{}}\PYG{n}{translate}\PYG{p}{(}\PYG{n}{x0x1}\PYG{p}{);}
\end{Verbatim}

\end{tcolorbox}
\caption{Compiling Quil kernels for the IBM back-end.}
\label{fig:compile}
\end{figure}

\texttt{CompositeInstructions} can be dynamic in that its children may not be known until runtime. To accommodate this, \texttt{CompositeInstruction} exposes an \texttt{expand()} method which takes a heterogeneous mapping of key-value pairs that seed the generation of children instructions. A prototypical example of this would be in programming recursive circuits such as the quantum Fourier transform, or a chemistry unitary coupled cluster circuit based on the number of qubits and electrons in the system. CompositeInstructions expose a \texttt{toGraph()} method call mapping the list of children instructions to a corresponding directed acyclic graph representation, useful in compilation and optimization routines. Finally, the \texttt{CompositeInstruction} can be evaluated at a set of concrete floating point parameters, each corresponding to one of the \texttt{CompositeInstruction}'s exposed variables.

To store and collect \texttt{CompositeInstructions}, XACC defines an \texttt{IR} interface which essentially aggregates constructed \texttt{CompositeInstructions}. In this way, the \texttt{IR} may be thought of as modeling a forest of \emph{n-ary} trees, where each tree can reference others in the same \texttt{IR} context. Both \texttt{IR} and \texttt{CompositeInstructions} can be persisted to a human-readable JSON string, as well as exposing capabilities to load itself from those persisted JSON strings. This enables caching of compilation results for reuse in later compilation contexts. As seen in Fig. \ref{fig:compile}, \texttt{IR} also exposes an API for querying or getting reference to compiled \texttt{CompositeInstructions}.

Complementing the \texttt{IR} interface, XACC defines an \texttt{IRTransformation} interface, which exposes a general \texttt{transform()} method taking \texttt{IR} in and outputting a new modified \texttt{IR} instance. This interface, and implementations of it, are critical for quantum compilation routines that seek to optimize quantum programs with regards to available back-end resources, enable error mitigation routines, or ensuring that quantum program connectivity is amenable for execution on the specific back-end connectivity.

To illustrate the \texttt{Instruction} and \texttt{CompositeInstruction} concepts, see the quantum kernel construction in Fig.~\ref{fig:compile}. The Quil compiler implementation in this example will parse the input source string, instantiate concrete low-level instructions for each line, and add them to a \texttt{CompositeInstruction}. Here each concrete instruction is provided to XACC as a subclass of the \texttt{Instruction} interface. The process begins by creating an \texttt{X} instruction, and setting its \texttt{bits} vector to \texttt{\{0\}}. Next the compiler creates a parameterized \texttt{Ry} gate, with a string \texttt{InstructionParameter} \texttt{x}. Finally a \texttt{CNOT} instruction is created with its \texttt{bits} as \texttt{\{1,0\}}. These will be added to a \texttt{CompositeInstruction} (with \texttt{variables = \{x\}}) and stored in the to-be-returned \texttt{IR} instance. Once a \texttt{CompositeInstruction} is created and stored, it can be referenced by other \texttt{CompositeInstructions}. This is how the second kernel is able to make reference to \texttt{ansatz} and add a couple of the \texttt{X} basis measurements.

\begin{figure}[!b]
\centering
\begin{tcolorbox}[colback=white]
\begin{Verbatim}[commandchars=\\\{\}]
\PYG{k}{auto} \PYG{n}{provider} \PYG{o}{=} \PYG{n}{xacc}\PYG{o}{::}\PYG{n}{getIRProvider}\PYG{p}{(}\PYG{l+s}{\PYGZdq{}quantum\PYGZdq{}}\PYG{p}{);}
\PYG{k}{auto} \PYG{n}{kernel} \PYG{o}{=}
  \PYG{n}{provider}\PYG{o}{\PYGZhy{}\PYGZgt{}}\PYG{n}{createComposite}\PYG{p}{(}\PYG{l+s}{\PYGZdq{}foo\PYGZdq{}}\PYG{p}{,\PYGZob{}}\PYG{l+s}{\PYGZdq{}theta\PYGZdq{}}\PYG{p}{\PYGZcb{});}
\PYG{k}{auto} \PYG{n}{x} \PYG{o}{=} \PYG{n}{provider}\PYG{o}{\PYGZhy{}\PYGZgt{}}\PYG{n}{createInstruction}\PYG{p}{(}\PYG{l+s}{\PYGZdq{}X\PYGZdq{}}\PYG{p}{,} \PYG{p}{\PYGZob{}}\PYG{l+m+mi}{0}\PYG{p}{\PYGZcb{});}
\PYG{k}{auto} \PYG{n}{ry} \PYG{o}{=}
  \PYG{n}{provider}\PYG{o}{\PYGZhy{}\PYGZgt{}}\PYG{n}{createInstruction}\PYG{p}{(}\PYG{l+s}{\PYGZdq{}Ry\PYGZdq{}}\PYG{p}{,} \PYG{p}{\PYGZob{}}\PYG{l+s}{\PYGZdq{}theta\PYGZdq{}}\PYG{p}{\PYGZcb{});}
\PYG{k}{auto} \PYG{n}{cx} \PYG{o}{=}
  \PYG{n}{provider}\PYG{o}{\PYGZhy{}\PYGZgt{}}\PYG{n}{createInstruction}\PYG{p}{(}\PYG{l+s}{\PYGZdq{}CX\PYGZdq{}}\PYG{p}{,} \PYG{p}{\PYGZob{}}\PYG{l+m+mi}{1}\PYG{p}{,}\PYG{l+m+mi}{0}\PYG{p}{\PYGZcb{});}
\PYG{n}{kernel}\PYG{o}{\PYGZhy{}\PYGZgt{}}\PYG{n}{addInstructions}\PYG{p}{(\PYGZob{}}\PYG{n}{x}\PYG{p}{,}\PYG{n}{ry}\PYG{p}{,}\PYG{n}{cx}\PYG{p}{\PYGZcb{});}
\end{Verbatim}
\end{tcolorbox}
\caption{Leveraging the \texttt{IRProvider} to create components of the \texttt{IR} without knowledge of underlying types.}
\label{fig:irprovider}
\end{figure}

A common pattern accompanying a tree of similar types is the visitor pattern, which enables type-specific operations to be added to a tree of common types at runtime via double dispatch. Essentially, the visitor pattern provides a mechanism for separating operations on a set of common objects from the structure of the objects themselves. XACC provides an implementation of the visitor pattern for \texttt{Instruction}, and in doing so provides an extensible mechanism for operations on the \texttt{IR}. Specifically, all XACC Instructions expose an \texttt{accept()} method that takes as input an \texttt{InstructionVisitor}. These visitors define a set of \texttt{visit} operations for each concrete \texttt{Instruction} sub-type it would like to visit, and a null implementation for all those it does not implement. Each \texttt{Instruction} subclass implements the \texttt{accept} call which invokes \texttt{visit} on the input \texttt{InstructionVisitor} giving itself as the argument to the \texttt{visit} call, thereby giving the visitor type information about the \texttt{Instruction}. This pattern enables a number of operations on the \texttt{IR} tree that is pertinent to quantum compilation and execution. First, specific back-ends can implement the \texttt{InstructionVisitor} and its desired \texttt{visit} methods to map \texttt{IR} to the corresponding native assembly. Moreover, simulation back-ends can implement the visitor to execute specific simulation methods, e.g. performing unitary operations on a state vector simulator. The extensibility of this pattern gives XACC developers maximal flexibility to design operations on the \texttt{IR} tree.
\begin{figure}[!b]
\centering
\includegraphics[width=\columnwidth]{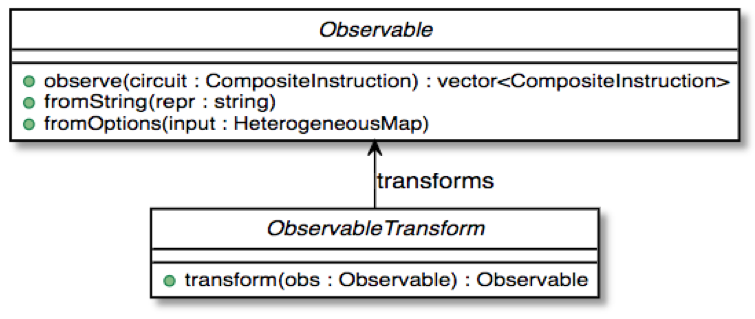}
\caption{The Observable and ObservableTransform interfaces. }
\label{fig:obs}
\end{figure}

Finally, the XACC IR package exposes a factory for the general creation of \texttt{Instructions}, \texttt{Composites}, and \texttt{IR}. The \texttt{IRProvider} interface exposes \texttt{createInstruction}, \texttt{createComposite}, and \texttt{createIR} methods that create subtypes of these interfaces. The code snippet in Fig. \ref{fig:irprovider} demonstrates how the \texttt{IRProvider} decouples programmers from underlying concrete type information for instructions, composite instructions, and IR.

\subsection{Observable and Observable Transform}
\begin{figure}[!t]
\centering
\begin{tcolorbox}[colback=white]
\begin{Verbatim}[commandchars=\\\{\}]
\PYG{c+c1}{// Observable from string}
\PYG{k}{auto} \PYG{n}{x0x1} \PYG{o}{=} \PYG{n}{xacc}\PYG{o}{::}\PYG{n}{getObservable}\PYG{p}{(}\PYG{l+s}{\PYGZdq{}pauli\PYGZdq{}}\PYG{p}{);}
\PYG{n}{x0x1}\PYG{o}{\PYGZhy{}\PYGZgt{}}\PYG{n}{fromString}\PYG{p}{(}\PYG{l+s}{\PYGZdq{}X0 X1\PYGZdq{}}\PYG{p}{);}

\PYG{c+c1}{// Measure, adds hadamards on 0,1}
\PYG{c+c1}{// and measure instructions}
\PYG{k}{auto} \PYG{n}{measured\PYGZus{}circuit} \PYG{o}{=}
        \PYG{n}{x0x1}\PYG{o}{\PYGZhy{}\PYGZgt{}}\PYG{n}{observe}\PYG{p}{(}\PYG{n}{circuit}\PYG{p}{)[}\PYG{l+m+mi}{0}\PYG{p}{];}

\PYG{c+c1}{// Observable from options}
\PYG{k}{auto} \PYG{n}{h2} \PYG{o}{=} \PYG{n}{xacc}\PYG{o}{::}\PYG{n}{getObservable}\PYG{p}{(}\PYG{l+s}{\PYGZdq{}chemistry\PYGZdq{}}\PYG{p}{);}
\PYG{n}{h2}\PYG{o}{\PYGZhy{}\PYGZgt{}}\PYG{n}{fromOptions}\PYG{p}{(\PYGZob{}}
    \PYG{p}{\PYGZob{}}\PYG{l+s}{\PYGZdq{}basis\PYGZdq{}}\PYG{p}{,} \PYG{l+s}{\PYGZdq{}sto\PYGZhy{}3g\PYGZdq{}}\PYG{p}{\PYGZcb{},}
    \PYG{p}{\PYGZob{}}\PYG{l+s}{\PYGZdq{}geometry\PYGZdq{}}\PYG{p}{,} \PYG{l+s+sa}{R}\PYG{l+s}{\PYGZdq{}}\PYG{l+s+dl}{(}
\PYG{l+s}{                    2}
\PYG{l+s}{                    H 0.0 0.0 0.0}
\PYG{l+s}{                    H 0.0 0.0 0.75}
\PYG{l+s}{                    }\PYG{l+s+dl}{)}\PYG{l+s}{\PYGZdq{}}
    \PYG{p}{\PYGZcb{}\PYGZcb{});}
\end{Verbatim}

\end{tcolorbox}
\caption{Demonstration of how programmers may create XACC \texttt{Observables}. Observables can be defined through appropriate subclasses of the \texttt{Observable}. Here we show general Pauli \texttt{Observables}, as well as a more complex chemistry \texttt{observable}.}
\label{fig:obs_code}
\end{figure}
The observable concept and transformations on observables have been proposed in the QCOR language specification \cite{qcor_spec}. In part, this specification describes a quantum observable as a rule over a range of qubits defining the measurement bases to be appended after an input circuit. It stipulates that implementations of the observable concept expose an \texttt{observe()} method which takes an unmeasured circuit or program and returns a list of (partially) measured circuits based on the observable's algebraic structure. The specification leaves the exact description of the circuit or kernel data structures to specification implementors. Moreover, the specification mandates that the observable concept expose methods for constructing the observable from both a string-like representation and a heterogeneous map of input parameters. Transformations on observables simply map one observable to another, and implementations are free to propose any number of transformations (like Jordan-Wigner, Bravyi-Kitaev, etc.).

We have implemented this concept in XACC via the definition of an \texttt{Observable} interface. XACC implements the \texttt{observe} method to take as input an unmeasured \texttt{CompositeInstruction} and return a list of measured \texttt{CompositeInstruction}s. We also define the \texttt{ObservableTransform} to take an Observable and map it to a new Observable. This transformation interface enables an extensible mechanism for contributing fermion-spin mappings (Jordan-Wigner, Bravyi-Kitaev, etc.) and operator symmetry reductions (e.g. qubit tapering \cite{tapering}). The class structure for these interfaces is shown in Fig. \ref{fig:obs}. Fig. \ref{fig:obs_code} demonstrates how Observables may be created in code.

\subsection{Accelerator}
\label{sec:acc}
\begin{figure}[!b]
\centering
\includegraphics[width=\columnwidth]{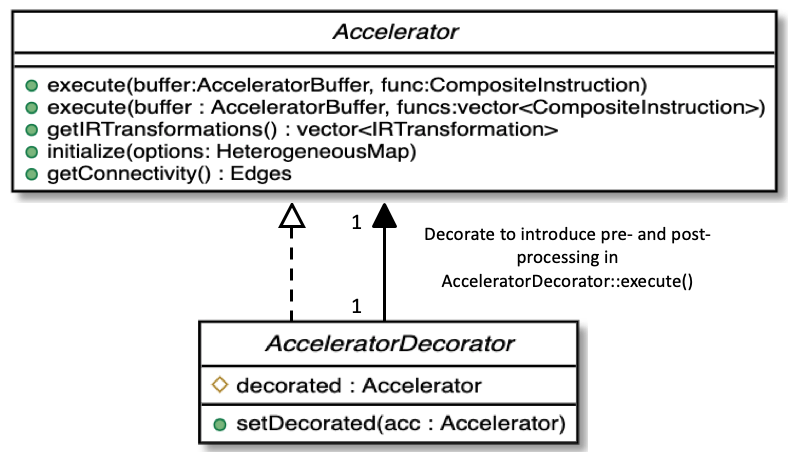}
% \missingfigure[figwidth=\columnwidth]{obs}
\caption{The class architecture for the \texttt{Accelerator} and \texttt{AcceleratorDecorator}. The \texttt{Accelerator} provides an extension point for general quantum back-ends, while the decorator enables general pre- and post-processing around quantum back-end execution.}
\label{fig:acc}
\end{figure}
The primary interface put forward by the XACC back-end layer is the \texttt{Accelerator}. This concept is intended to be implemented for physical and virtual quantum computing back-ends that can be local or remote. Accelerators expose an \texttt{execute()} method that takes as input an \texttt{AcceleratorBuffer} instance and the \texttt{CompositeInstruction} containing the compiled representation of the quantum program. Accelerators are free to implement this as necessary for executing the program on their representative hardware or simulator, but must persist all measurement results to the provided buffer instance. Implementations can also take advantage of the buffer heterogeneous metadata member to persist any other pertinent information (such as readout error probabilities, $T_1$ and $T_2$ times, or expectation values). The \texttt{execute} method is overloaded to enable the execution of multiple \texttt{CompositeInstructions} at once to improve on the number of remote process invocations required to execute a given problem.

\texttt{Accelerators} can be initialized with a heterogeneous map of options, a useful feature for tweaking the number of shots, the desired remote back-end, and any other execution parameters. \texttt{Accelerators} can return a list of hardware-dependent \texttt{IRTransformations} that will be run on the incoming \texttt{CompositeInstruction} to ensure that the program is amenable for execution on the accelerator. \texttt{Accelerators} also expose the underlying qubit connectivity via the \texttt{getConnectivity()} method. This can be used in compilation routines or other \texttt{IRTransformation} implementations.
\begin{figure}[!t]
\centering
\begin{tcolorbox}[colback=white]
\begin{Verbatim}[commandchars=\\\{\}]
\PYG{c+c1}{// Get a concrete accelerator}
\PYG{k}{auto} \PYG{n}{ibm} \PYG{o}{=} \PYG{n}{xacc}\PYG{o}{::}\PYG{n}{getAccelerator}\PYG{p}{(}\PYG{l+s}{\PYGZdq{}ibm:tokyo\PYGZdq{}}\PYG{p}{,}
                    \PYG{p}{\PYGZob{}\PYGZob{}}\PYG{l+s}{\PYGZdq{}shots\PYGZdq{}}\PYG{p}{,}\PYG{l+m+mi}{8192}\PYG{p}{\PYGZcb{}\PYGZcb{});}

\PYG{c+c1}{// get an accelerator decorator}
\PYG{k}{auto} \PYG{n}{ro\PYGZus{}decorator} \PYG{o}{=}
    \PYG{n}{xacc}\PYG{o}{::}\PYG{n}{getAcceleratorDecorator}\PYG{p}{(}\PYG{l+s}{\PYGZdq{}ro\PYGZhy{}error\PYGZdq{}}\PYG{p}{);}

\PYG{c+c1}{// decorate}
\PYG{n}{ro\PYGZus{}decorator}\PYG{o}{\PYGZhy{}\PYGZgt{}}\PYG{n}{decorate}\PYG{p}{(}\PYG{n}{ibm}\PYG{p}{);}

\PYG{c+c1}{// execute just like you}
\PYG{c+c1}{// would any accelerator}
\PYG{n}{ro\PYGZus{}decorator}\PYG{o}{\PYGZhy{}\PYGZgt{}}\PYG{n}{execute}\PYG{p}{(}\PYG{n}{buffer}\PYG{p}{,} \PYG{n}{circuit}\PYG{p}{)}

\PYG{c+c1}{// buffer has readout\PYGZhy{}error}
\PYG{c+c1}{// corrected results...}
\PYG{k}{auto} \PYG{n}{ro\PYGZus{}corrected} \PYG{o}{=}
    \PYG{n}{buffer}\PYG{o}{\PYGZhy{}\PYGZgt{}}\PYG{n}{getInformation}\PYG{p}{(}\PYG{l+s}{\PYGZdq{}exp\PYGZhy{}val\PYGZdq{}}\PYG{p}{);}
\end{Verbatim}

\end{tcolorbox}
\caption{A code snippet demonstrating how one might leverage \texttt{Accelerators} and \texttt{AcceleratorDecorators} to automate certain error mitigation strategies.}
\label{fig:accd_code}
\end{figure}

XACC defines a decorator pattern on the \texttt{Accelerator} interface that promotes extensible pre- and post-processing of \texttt{Accelerator} execution input and results. The \texttt{AcceleratorDecorator} interface is an \texttt{Accelerator} sub-type but also delegates to a concrete \texttt{Accelerator} (see Fig.~\ref{fig:acc}). It implements \texttt{Accelerator::execute()} to insert pre-processing of execution input, execution on the delegated \texttt{Accelerator}, and post-processing of measurement results and metadata. This design is particularly useful in near-term quantum computation as it enables an extension point for general error mitigation strategies. For example, one could implement automated qubit measurement readout-error mitigation by implementing the \texttt{AcceleratorDecorator} to post-process computed expectation values according to shift rules based on known readout error probabilites \cite{Dumitrescu2018}. The snippet in Fig. \ref{fig:accd_code} demonstrates how one might leverage this capability to automate certain error mitigation strategies. It is useful to note that this pattern enables a chain of decorators, with the final decorator delegating to a concrete \texttt{Accelerator}, thereby enabling a composition of different pre- and post-processing strategies (multiple error mitigation strategies).

\subsection{Algorithms}
XACC defines a general \texttt{Algorithm} interface to enable the injection of pre-defined quantum-classical algorithms and therefore relieve programmers of the burden of having to program them from scratch. XACC defines an \texttt{Algorithm} as a concept or abstraction that takes as input a heterogeneous map of input data, and exposes a mechanism for execution with a provided \texttt{AcceleratorBuffer}. Examples of algorithm implementations provided by XACC are the variational quantum eigensolver, reduced density matrix element generation for quantum chemistry calculations, and a data-driven circuit learning algorithm for machine learning tasks that leverage quantum back-ends. Sec.~\ref{sec:demo} demonstrates the utility of this \texttt{Algorithm} interface.
\subsection{Optimizer}
\begin{figure}[!b]
\centering
\begin{tcolorbox}[colback=white]
\begin{Verbatim}[commandchars=\\\{\}]
\PYG{k}{auto} \PYG{n}{optimizer} \PYG{o}{=}
      \PYG{n}{xacc}\PYG{o}{::}\PYG{n}{getOptimizer}\PYG{p}{(}\PYG{l+s}{\PYGZdq{}nlopt\PYGZdq{}}\PYG{p}{);}

\PYG{n}{OptFunction} \PYG{n+nf}{f}\PYG{p}{(}
    \PYG{p}{[](}\PYG{k}{const} \PYG{n}{std}\PYG{o}{::}\PYG{n}{vector}\PYG{o}{\PYGZlt{}}\PYG{k+kt}{double}\PYG{o}{\PYGZgt{}} \PYG{o}{\PYGZam{}}\PYG{n}{x}\PYG{p}{,}
        \PYG{n}{std}\PYG{o}{::}\PYG{n}{vector}\PYG{o}{\PYGZlt{}}\PYG{k+kt}{double}\PYG{o}{\PYGZgt{}} \PYG{o}{\PYGZam{}}\PYG{n}{grad}\PYG{p}{)} \PYG{p}{\PYGZob{}}

    \PYG{c+c1}{// any preprocessing needed before}
    \PYG{c+c1}{// back\PYGZhy{}end execution}

    \PYG{c+c1}{// Execute on back\PYGZhy{}end Accelerator}
    \PYG{n}{qpu}\PYG{o}{\PYGZhy{}\PYGZgt{}}\PYG{n}{execute}\PYG{p}{(}\PYG{n}{buffer}\PYG{p}{,} \PYG{n}{circuits}\PYG{p}{);}

    \PYG{c+c1}{// post\PYGZhy{}processing on results}
    \PYG{c+c1}{// compute f\PYGZus{}val scalar, compute}
    \PYG{c+c1}{// gradient if necessary}

    \PYG{k}{return} \PYG{n}{f\PYGZus{}val}\PYG{p}{;}
    \PYG{p}{\PYGZcb{},}
    \PYG{n}{n\PYGZus{}params}\PYG{p}{);}

\PYG{n}{optimizer}\PYG{o}{\PYGZhy{}\PYGZgt{}}\PYG{n}{setOptions}\PYG{p}{(}
    \PYG{n}{HeterogeneousMap}\PYG{p}{\PYGZob{}}
      \PYG{n}{std}\PYG{o}{::}\PYG{n}{make\PYGZus{}pair}\PYG{p}{(}\PYG{l+s}{\PYGZdq{}nlopt\PYGZhy{}maxeval\PYGZdq{}}\PYG{p}{,} \PYG{l+m+mi}{200}\PYG{p}{),}
      \PYG{n}{std}\PYG{o}{::}\PYG{n}{make\PYGZus{}pair}\PYG{p}{(}\PYG{l+s}{\PYGZdq{}nlopt\PYGZhy{}optimizer\PYGZdq{}}\PYG{p}{,}
                      \PYG{l+s}{\PYGZdq{}l\PYGZhy{}bfgs\PYGZdq{}}\PYG{p}{)}
    \PYG{p}{\PYGZcb{});}

\PYG{k}{auto} \PYG{n}{result} \PYG{o}{=} \PYG{n}{optimizer}\PYG{o}{\PYGZhy{}\PYGZgt{}}\PYG{n}{optimize}\PYG{p}{(}\PYG{n}{f}\PYG{p}{);}
\end{Verbatim}

\end{tcolorbox}
\caption{A code snippet demonstrating how one might leverage \texttt{Optimizers}.}
\label{fig:opt_code}
\end{figure}
The QCOR specification puts forward a general optimizer concept \cite{qcor_spec}. This abstraction is meant to provide a general extension point for methods that optimize general multi-variate functions, and specifically for quantum computing, functions that internally make calls to a quantum back-end. XACC implements this concept via an \texttt{Optimizer} interface, which exposes an \texttt{optimize()} method that takes as input a general \texttt{OptFunction}. The \texttt{OptFunction} is a thin class that wraps a C\texttt{++} \texttt{std::function<double(const std::vector<double>, std::vector<double>\&)>} instance representing the function to be optimized. This function instance must take as its first argument a \texttt{const std::vector<double>} representing the current optimization iteration parameters, and as its second input a \texttt{std::vector<double>\&} representing the mutable function gradient with respect to the current iterate's parameters. It returns a double representing the evaluation of the function at the provided parameters. \texttt{Optimizer} implementations are free to ignore the second argument in derivative-free optimization routines. \texttt{Optimizers} also expose a \texttt{setOptions()} method that takes as input a heterogeneous map of input data, enabling programmers to configure the underlying optimization routines being leveraged.

XACC has concrete implementations of the \texttt{Optimizer} that delegate to the well-known \texttt{NLOpt} \cite{nlopt} and \texttt{mlpack} \cite{mlpack2018} libraries. The code snippet in Fig.~\ref{fig:opt_code} demonstrates the utility of this extension point.

\subsection{The XACC Service Registry}
At its core, XACC architecture puts forward interfaces for accelerators, accelerator decorators, instructions, composite instructions, IR transformations, compilers, algorithms, observables, and observable transforms. XACC sees these interfaces as extension points for the framework, with implementations serving as available services for the various aspects of the quantum programming and execution workflow. In an effort to ensure modularity and extensibility, XACC only handles service implementations as standard (shared) pointers to the interface being implemented. In this way, XACC never knows about underlying types, implying that a single XACC install can be built but the underlying functionality can change without a total re-build of the framework.
\begin{figure}[!t]
\centering
\begin{tcolorbox}[colback=white]
\begin{Verbatim}[commandchars=\\\{\}]
\PYG{c+c1}{// Create CppMicroServices Framework}
\PYG{k}{auto} \PYG{n}{framework} \PYG{o}{=}
    \PYG{n}{FrameworkFactory}\PYG{p}{().}\PYG{n}{NewFramework}\PYG{p}{();}

\PYG{c+c1}{// Initialize the Framework}
\PYG{n}{framework}\PYG{p}{.}\PYG{n}{Init}\PYG{p}{();}

\PYG{c+c1}{// Get the Bundle Context}
\PYG{k}{auto} \PYG{n}{context} \PYG{o}{=} \PYG{n}{framework}\PYG{p}{.}\PYG{n}{GetBundleContext}\PYG{p}{();}

\PYG{c+c1}{// Load and install plugins}
\PYG{k}{auto} \PYG{n}{plugins} \PYG{o}{=}
    \PYG{n}{loadPluginLibraries}\PYG{p}{(}\PYG{l+s}{\PYGZdq{}\PYGZti{}/.xacc/plugins\PYGZdq{}}\PYG{p}{);}
\PYG{k}{for} \PYG{p}{(}\PYG{k}{auto}\PYG{o}{\PYGZam{}} \PYG{n+nl}{plugin} \PYG{p}{:} \PYG{n}{plugins}\PYG{p}{)} \PYG{p}{\PYGZob{}}
  \PYG{n}{context}\PYG{p}{.}\PYG{n}{InstallBundles}\PYG{p}{(}\PYG{n}{plugin}\PYG{p}{);}
\PYG{p}{\PYGZcb{}}

\PYG{c+c1}{// Start the framework.}
\PYG{n}{framework}\PYG{p}{.}\PYG{n}{Start}\PYG{p}{();}

\PYG{c+c1}{// Get all installed bundles and install them}
\PYG{k}{auto} \PYG{n}{bundles} \PYG{o}{=} \PYG{n}{context}\PYG{p}{.}\PYG{n}{GetBundles}\PYG{p}{();}
\PYG{k}{for} \PYG{p}{(}\PYG{k}{auto} \PYG{n+nl}{b} \PYG{p}{:} \PYG{n}{bundles}\PYG{p}{)} \PYG{p}{\PYGZob{}}
    \PYG{n}{b}\PYG{p}{.}\PYG{n}{Start}\PYG{p}{();}
\PYG{p}{\PYGZcb{}}

\PYG{c+c1}{// ... Later requesting a certain Service}

\PYG{n}{std}\PYG{o}{::}\PYG{n}{string} \PYG{n}{name} \PYG{o}{=} \PYG{l+s}{\PYGZdq{}openqasm\PYGZdq{}}\PYG{p}{;}
\PYG{n}{xacc}\PYG{o}{::}\PYG{n}{Compiler} \PYG{n}{compiler}\PYG{p}{;}
\PYG{k}{auto} \PYG{n}{compilers} \PYG{o}{=}
    \PYG{n}{context}\PYG{p}{.}\PYG{n}{GetServiceReferences}\PYG{o}{\PYGZlt{}}\PYG{n}{Compiler}\PYG{o}{\PYGZgt{}}\PYG{p}{();}
\PYG{k}{for} \PYG{p}{(}\PYG{k}{auto}\PYG{o}{\PYGZam{}} \PYG{n+nl}{s} \PYG{p}{:} \PYG{n}{compilers}\PYG{p}{)} \PYG{p}{\PYGZob{}}
  \PYG{k}{auto} \PYG{n}{service} \PYG{o}{=} \PYG{n}{context}\PYG{p}{.}\PYG{n}{GetService}\PYG{p}{(}\PYG{n}{s}\PYG{p}{);}
  \PYG{k}{if} \PYG{p}{(}\PYG{n}{service} \PYG{o}{\PYGZam{}\PYGZam{}}
        \PYG{n}{service}\PYG{o}{\PYGZhy{}\PYGZgt{}}\PYG{n}{name}\PYG{p}{()} \PYG{o}{==} \PYG{n}{name}\PYG{p}{)} \PYG{p}{\PYGZob{}}
      \PYG{n}{compiler} \PYG{o}{=} \PYG{n}{service}\PYG{p}{;}
      \PYG{k}{break}\PYG{p}{;}
  \PYG{p}{\PYGZcb{}}
\PYG{p}{\PYGZcb{}}
\end{Verbatim}
\end{tcolorbox}
\caption{Example code snippet demonstrating the CppMicroServices OSGi implementation and its utility to XACC. The CppMicroServices framework is started, bundles loaded as shared libraries, and services are requested, all at runtime without direct dependence on third-party contributions or framework re-builds.}
\label{fig:code}
\end{figure}

To enable this, XACC requires a robust factory pattern that promotes this sort of interface-based programming. This is common in the Java, where major application frameworks provide modularity and extensibility through interface registration, dynamic loading, and reflection. To promote this in a standard way, the open-source community put forward the Open Services Gateway Initiative (OSGi) \cite{osgi}, which defines several abstractions common to modular frameworks with runtime-extensibility. The OSGi specification defines the concept of a \emph{bundle} containing service (interface) implementations that are declared to the framework at runtime via standard metadata files. These services are loaded and registered with a core framework instance, thus enabling clients to request an instance of an interface implementation given by a specified string name. This technology underlies most graphical integrated development environments (IDE), and enables integration of programming tools across a number of languages.

XACC picks up on these concepts, and leverages a C\texttt{++} implementation of the OSGi specification called \texttt{CppMicroServices} \cite{cppmicroservices}. This framework enables runtime registration of interface implementations against their interface type and unique string identifiers. These implementations are provided to the framework as plugins, or code that sits outside the core XACC context, but provides an implementation of a core XACC interface, and can be added or removed without affecting the core framework. These plugins are built and installed as standard shared libraries, and dropped into a specific folder that XACC defines. At runtime, XACC loads the plugins (shared libraries) into the core CppMicroServices framework and all available interface implementations are registered. Clients can then request specific services by providing the interface type and the name of the service. The code in Fig. \ref{fig:code} demonstrates this process. First, a CppMicroServices framework is created and initialized, then all plugin libraries are loaded and all bundles are started (thus registering all service implementations). Later, clients can request a service (here \texttt{Compiler}) of a specified unique name (\texttt{openqasm}).

To handle all of this, XACC leverages a class called the \texttt{ServiceRegistry}, which exposes templated methods to retrieve implementations corresponding to a certain interface type and the unique string name. This class exposes a \texttt{getService<TYPE>(name:string):TYPE} method allowing programmers to request a certain XACC interface implementation at the given name. The previously demonstrated code snippets in this work reference API calls like \texttt{xacc::getAccelerator()}, \texttt{xacc::getCompiler()}, etc. These are just wrappers around calls to the \texttt{ServiceRegistry::getService<T>(name:string)}, which returns the appropriate instance of type \texttt{T}.

\subsection{Heterogeneous Map}
\label{sec:utils}
A crucial utility data structure that has been mentioned throughout the previous discussion and code snippets, but not yet elaborated on, is the XACC \texttt{HeterogeneousMap}. This abstraction is described in the QCOR specification \cite{qcor_spec} as a mechanism or data structure for mapping string keys to any value type. This type of construct is commonplace in Python (\texttt{dict} is a heterogeneous mapping), but is more difficult to implement in a system-level, statically typed language like C\texttt{++}. The standard library implements a map data structure, but it must be instantiated with the concrete key and value types. In \texttt{C++17}, there are two ways how this can be achieved. One way to achieve this heterogeneous value-type behavior would be to use type erasure via \texttt{C++17} \texttt{std::any<T>}. Another way would be to create a standard map with string keys and variant-like value (\texttt{std::variant} in C\texttt{++17}). The XACC \texttt{HeterogeneousMap} leverages the \texttt{C++14} standard, specifically static template class members, in order to achieve heterogeneous value types, thus enabling the data structure and corresponding API demonstrated in Fig. \ref{fig:hmap}. %We implemented this class to contain a static templated \texttt{unordered_map} member mapping a \texttt{HeterogeneousMap*} to a \texttt{map} of \texttt{string}s to instances of the template type. Insertion corresponds to correctly creating or reusing the correct \texttt{map} keyed by \texttt{this} \texttt{HeterogeneousMap}.
\begin{figure}[!t]
\centering
\begin{tcolorbox}[colback=white]
\begin{Verbatim}[commandchars=\\\{\}]
\PYG{c+c1}{// Create a map and use insert()}
\PYG{c+c1}{// method leveraging template}
\PYG{c+c1}{// type deduction}
\PYG{n}{HeterogeneousMap} \PYG{n}{m}\PYG{p}{;}
\PYG{n}{m}\PYG{p}{.}\PYG{n}{insert}\PYG{p}{(}\PYG{l+s}{\PYGZdq{}int\PYGZhy{}key\PYGZdq{}}\PYG{p}{,} \PYG{l+m+mi}{1}\PYG{p}{);}
\PYG{n}{m}\PYG{p}{.}\PYG{n}{insert}\PYG{p}{(}\PYG{l+s}{\PYGZdq{}double\PYGZhy{}key\PYGZdq{}}\PYG{p}{,} \PYG{l+m+mf}{2.0}\PYG{p}{);}
\PYG{n}{m}\PYG{p}{.}\PYG{n}{insert}\PYG{p}{(}\PYG{l+s}{\PYGZdq{}vector\PYGZhy{}key\PYGZdq{}}\PYG{p}{,}
        \PYG{n}{std}\PYG{o}{::}\PYG{n}{vector}\PYG{o}{\PYGZlt{}}\PYG{k+kt}{double}\PYG{o}{\PYGZgt{}}\PYG{p}{\PYGZob{}}\PYG{l+m+mf}{1.0}\PYG{p}{,}\PYG{l+m+mf}{2.0}\PYG{p}{\PYGZcb{});}
\PYG{n}{m}\PYG{p}{.}\PYG{n}{insert}\PYG{p}{(}\PYG{l+s}{\PYGZdq{}string\PYGZhy{}key\PYGZdq{}}\PYG{p}{,} \PYG{n}{std}\PYG{o}{::}\PYG{n}{string}\PYG{p}{(}\PYG{l+s}{\PYGZdq{}hello\PYGZdq{}}\PYG{p}{));}

\PYG{c+c1}{// Create from initializer list}
\PYG{n}{HeterogeneousMap} \PYG{n}{m2}\PYG{p}{\PYGZob{}}
        \PYG{n}{std}\PYG{o}{::}\PYG{n}{make\PYGZus{}pair}\PYG{p}{(}\PYG{l+s}{\PYGZdq{}int\PYGZhy{}key\PYGZdq{}}\PYG{p}{,}\PYG{l+m+mi}{1}\PYG{p}{),}
        \PYG{n}{std}\PYG{o}{::}\PYG{n}{make\PYGZus{}pair}\PYG{p}{(}\PYG{l+s}{\PYGZdq{}double\PYGZhy{}key\PYGZdq{}}\PYG{p}{,}\PYG{l+m+mf}{2.0}\PYG{p}{)}
\PYG{p}{\PYGZcb{};}
\end{Verbatim}

\end{tcolorbox}
\caption{Demonstration of the XACC \texttt{HeterogeneousMap}. This utility class is used throughout XACC as a mechanism for handling heterogeneous, problem- and back-end-specific information.}
\label{fig:hmap}
\end{figure}

\subsection{XACC Framework API}
\label{sec:api}
\begin{figure}[!b]
\centering
\includegraphics[width=\columnwidth]{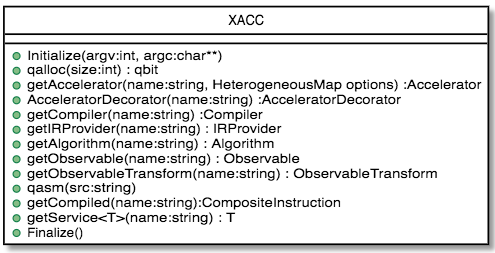}
\caption{The public XACC API provides programmers with an easy mechanism for interacting with the core XACC service registry, as well as allocating \texttt{AcceleratorBuffers} and compiling quantum kernels.}
\label{fig:xacc_api}
\end{figure}
The public XACC API has been demonstrated (in part) throughout the code snippets in the previous sections. Here we elaborate on the calls exposed in the API and detail the functionality each provides. This API is the primary interface clients/programmers leverage when interacting with XACC and its underlying service framework. The API is demonstrated graphically in Fig. \ref{fig:xacc_api}.

All applications leveraging the XACC framework must initialize the framework before invocation of any XACC API calls. XACC provides an \texttt{Initialize()} function for this, which takes as input the standard \texttt{argc}, \texttt{argv} command line argument variables. The primary function of this method is to parse and analyze any application command line options, and initialize the underlying XACC CppMicroServices plugin framework. This involves the searching, loading, and registering of shared libraries exposing core XACC service implementations. A corresponding \texttt{Finalize} API call is provided for invocation at the end of XACC usage that tears down the service registry and cleans up allocated memory.

The next API call that programmers will interact with is an allocator of quantum memory. In a similar vein as the C \texttt{malloc} call, XACC exposes a \texttt{qalloc} API call that allocates an instance of the \texttt{AcceleratorBuffer}. Note that at this level, XACC defines a \texttt{typedef} for \texttt{AcceleratorBuffer} called \texttt{qbit}. The primary purpose for this \texttt{typedef} is for readability, and it is leveraged in the QCOR single-source quantum-classical C\texttt{++} compiler \cite{qcor_in_prep}. \texttt{qbit} and \texttt{AcceleratorBuffer} can be used interchangeably.

XACC exposes public methods for getting references to implementations of core interfaces, such as accelerator, compiler, IR provider, algorithm, IR transformation, and observable implementations. Each implementation exposes a unique name (\texttt{ibm} for the IBM Accelerator, for instance), so each of these API calls (\texttt{getAccelerator}, \texttt{getCompiler}, etc.) takes as input the name of the service implementation, and returns a shared pointer to that instantiated service. Programmers can also directly leverage a templated \texttt{getService<T>(name:string) : T} call to instantiate a service of type \texttt{T} with the given string name.

\begin{figure}[!t]
\centering
\begin{tcolorbox}[colback=white]
\begin{Verbatim}[commandchars=\\\{\}]
\PYG{c+c1}{// JIT map XASM Ansatz to IR}
\PYG{n}{xacc}\PYG{o}{::}\PYG{n}{qasm}\PYG{p}{(}\PYG{l+s+sa}{R}\PYG{l+s}{\PYGZdq{}}\PYG{l+s+dl}{(}
\PYG{l+s}{.compiler xasm}
\PYG{l+s}{.circuit deuteron\PYGZus{}ansatz}
\PYG{l+s}{.parameters theta}
\PYG{l+s}{.qbit q}
\PYG{l+s}{X(q[0]);}
\PYG{l+s}{Ry(q[1], theta);}
\PYG{l+s}{CNOT(q[1],q[0]);}
\PYG{l+s+dl}{)}\PYG{l+s}{\PYGZdq{}}\PYG{p}{);}
\PYG{k}{auto} \PYG{n}{ansatz} \PYG{o}{=}
    \PYG{n}{xacc}\PYG{o}{::}\PYG{n}{getCompiled}\PYG{p}{(}\PYG{l+s}{\PYGZdq{}deuteron\PYGZus{}ansatz\PYGZdq{}}\PYG{p}{);}

\PYG{c+c1}{// Quil example, multiple kernels}
\PYG{n}{xacc}\PYG{o}{::}\PYG{n}{qasm}\PYG{p}{(}\PYG{l+s+sa}{R}\PYG{l+s}{\PYGZdq{}}\PYG{l+s+dl}{(}\PYG{l+s}{.compiler quil}
\PYG{l+s}{.circuit ansatz}
\PYG{l+s}{.parameters theta, phi}
\PYG{l+s}{X 0}
\PYG{l+s}{H 2}
\PYG{l+s}{CNOT 2 1}
\PYG{l+s}{CNOT 0 1}
\PYG{l+s}{Rz(theta) 0}
\PYG{l+s}{Ry(phi) 1}
\PYG{l+s}{H 0}
\PYG{l+s}{.circuit x0x1}
\PYG{l+s}{ansatz(theta, phi)}
\PYG{l+s}{H 0}
\PYG{l+s}{H 1}
\PYG{l+s}{MEASURE 0 [0]}
\PYG{l+s}{MEASURE 1 [1]}
\PYG{l+s+dl}{)}\PYG{l+s}{\PYGZdq{}}\PYG{p}{);}
\PYG{k}{auto} \PYG{n}{x0x1} \PYG{o}{=} \PYG{n}{xacc}\PYG{o}{::}\PYG{n}{getCompiled}\PYG{p}{(}\PYG{l+s}{\PYGZdq{}x0x1\PYGZdq{}}\PYG{p}{);}
\end{Verbatim}

\end{tcolorbox}
\caption{Demonstration of the XACC \texttt{qasm()} call. Note multiple kernel strings can be compiled and reference each other.}
\label{fig:qasm}
\end{figure}

To improve programming efficiency, readability, and utility of the quantum kernel string compilation (like in Fig. \ref{fig:compile}), XACC exposes a \texttt{qasm()} function. This function takes as input an enhanced quantum kernel source string syntax and compiles it to XACC IR. This source string is \emph{enhanced} in that it requires that extra metadata be present in order to adequately compile the quantum code. Specifically, the source string must contain the following key words:
\begin{itemize}
    \item a single \texttt{.compiler NAME}, to indicate which XACC compiler implementation to use.
    \item one or many \texttt{.circuit NAME} calls to give the created CompositeInstruction (circuit) a name.
    \item one \texttt{.parameters PARAM\_NAME\_1, PARAM\_NAME\_2, \ldots, PARAM\_NAME\_N} for each parameterized circuit to tell the Compiler the names of the parameters.
    \item A \texttt{.qbit NAME} optional keyword can be provided when the source code itself makes reference to the \texttt{qbit} or \texttt{AcceleratorBuffer}
\end{itemize}
Running this command with the appropriately provided keywords will compile the source string to XACC IR and store it an internal compilation database (standard map of CompositeInstruction names to CompositeInstructions), and users can get reference to the individual CompositeInstructions via an exposed \texttt{getCompiled()} XACC API call. The code in Fig. \ref{fig:qasm} demonstrates how one would use \texttt{qasm()} and its overall utility.

\section{Interface Implementations for Quantum Computing}
\label{sec:inter}
Having introduced the core interfaces exposed by core XACC, we now turn to detailing concrete implementations of these interfaces for quantum computation. Specifically, we will discuss gate and annealing model implementations of the core XACC IR interfaces, Pauli and Fermion implementations of the \texttt{Observable} interface (and Jordan-Wigner, Bravyi-Kitaev implementations of \texttt{ObservableTransform}), \texttt{Compiler} implementations for standard low-level assembly languages (Quil, OpenQasm, XACC XASM) using Antlr \cite{antlr}, \texttt{Accelerators} for available QPUs and simulators, and \texttt{Algorithm} implementations. We note that for most use cases programmers do not need to know the concrete type of these implementations as instances are created and provided as pointers to the interface itself via the XACC public API. The implementations to be discussed here are available in the \texttt{xacc::quantum} package.

XACC defines \texttt{Gate} and \texttt{Circuit} classes implementing \texttt{Instruction} and \texttt{CompositeInstruction}, respectively. To hide this type information from programmers, XACC defines a \texttt{QuantumIRProvider} which implements the \texttt{IRProvider} interface for the creation of \texttt{Gates} and \texttt{Circuits} (returned as pointers to \texttt{Instruction} and \texttt{Circuit}). The \texttt{Gate} implements the aforementioned \texttt{Instruction} API and is further subclassed by types providing specific digital gate implementations, such as \texttt{Hadamard}, \texttt{CNOT}, \texttt{Rz}, etc. The \texttt{Circuit} class implements the aforementioned \texttt{CompositeInstruction} API to provide a container for \texttt{Instructions} while also remaining a valid \texttt{Instruction} itself. The \texttt{Circuit} may contain further \texttt{Circuits}, enabling the familiar \emph{n-ary} tree pattern. \texttt{Circuits} keep track of string variables that it and sub-\texttt{Instructions} depend on, and implements \texttt{CompositeInstruction::operator()()} to take concrete double values and evaluate all sub-\texttt{Instructions} dependent on a specific \texttt{Circuit} variables. A key aspect of \texttt{Circuits} is that they can be \emph{dynamic}, i.e. sub-types of \texttt{Circuit} can override the \texttt{CompositeInstruction::expand} method taking an \texttt{HeterogeneousMap} of input data and expanding itself with concrete \texttt{Gate} instructions. This is useful for dynamic, pre-defined composite instructions that can be programmed at compile-time but expanded at runtime. A few examples of this, which are implemented in XACC, are the following \texttt{Circuit} sub-types:
\begin{itemize}
    \item \texttt{range} Takes as input a \texttt{Gate} name and the range of qubits (given as a standard range \texttt{[start\_idx,end\_idx)}, or the total number of qubits). Based on this input it adds that \texttt{Gate} to all qubits in the range. Typical invocation using the XASM compiler looks like
    \texttt{range(buffer, \{\{"gate", "H"\}, \{"nq",4\}\});}.
    \item \texttt{qft} Takes as input the range of qubits (given as a standard range \texttt{[start\_idx,end\_idx)}, or the total number of qubits) upon which to append a quantum Fourier transform. Typical invocation using the XASM compiler looks like \texttt{qft(buffer, \{\{"nq", 4\}\});}.
    \item \texttt{exp\_i\_theta} Takes as input a representation of a spin or fermionic operator $H$ and provides a first order trotterization of $\exp(i \theta H)$. Typical invocation using the XASM compiler looks like \texttt{exp\_i\_theta(buffer, theta, \{\{"pauli", "X0 X1 - Y0 Y1"\}\});}.
\end{itemize}
An example of using these dynamic circuit generators with the \texttt{xacc::qasm()} call is shown in Fig. \ref{fig:qasm_generator}.
\begin{figure}[!t]
\centering
\begin{tcolorbox}[colback=white]
\begin{Verbatim}[commandchars=\\\{\}]
\PYG{c+c1}{// JIT map XASM Ansatz to IR}
\PYG{n}{xacc}\PYG{o}{::}\PYG{n}{qasm}\PYG{p}{(}\PYG{l+s+sa}{R}\PYG{l+s}{\PYGZdq{}}\PYG{l+s+dl}{(}
\PYG{l+s}{.compiler xasm}
\PYG{l+s}{.circuit deuteron\PYGZus{}ansatz}
\PYG{l+s}{.parameters x}
\PYG{l+s}{.qbit q}
\PYG{l+s}{for (int i = 0; i \PYGZlt{} 2; i++) \PYGZob{}}
\PYG{l+s}{  H(q[0]);}
\PYG{l+s}{\PYGZcb{}}
\PYG{l+s}{exp\PYGZus{}i\PYGZus{}theta(q, x, \PYGZob{}\PYGZob{}\PYGZdq{}pauli\PYGZdq{}, \PYGZdq{}X0 Y1 \PYGZhy{} Y0 X1\PYGZdq{}\PYGZcb{}\PYGZcb{});}
\PYG{l+s+dl}{)}\PYG{l+s}{\PYGZdq{}}\PYG{p}{);}
\PYG{k}{auto} \PYG{n}{ansatz} \PYG{o}{=}
    \PYG{n}{xacc}\PYG{o}{::}\PYG{n}{getCompiled}\PYG{p}{(}\PYG{l+s}{\PYGZdq{}deuteron\PYGZus{}ansatz\PYGZdq{}}\PYG{p}{);}
\end{Verbatim}
\end{tcolorbox}
\caption{Demonstration of the XACC XASM Compiler through the \texttt{qasm()} call.}
\label{fig:qasm_generator}
\end{figure}

\begin{figure}[!b]
\centering
\begin{tcolorbox}[colback=white]
\begin{Verbatim}[commandchars=\\\{\}]
\PYG{c+c1}{// Note Accelerators can be initialized}
\PYG{c+c1}{// with a HeterogeneousMap.}
\PYG{k}{auto} \PYG{n}{ibm} \PYG{o}{=} \PYG{n}{xacc}\PYG{o}{::}\PYG{n}{getAccelerator}\PYG{p}{(}\PYG{l+s}{\PYGZdq{}ibm\PYGZdq{}}\PYG{p}{,}
            \PYG{p}{\PYGZob{}\PYGZob{}}\PYG{l+s}{\PYGZdq{}shots\PYGZdq{}}\PYG{p}{,} \PYG{l+m+mi}{8192}\PYG{p}{\PYGZcb{},}
            \PYG{p}{\PYGZob{}}\PYG{l+s}{\PYGZdq{}backend\PYGZdq{}}\PYG{p}{,} \PYG{l+s}{\PYGZdq{}ibmq\PYGZus{}poughkeepsie\PYGZdq{}}\PYG{p}{\PYGZcb{}\PYGZcb{});}
\PYG{n}{ibm}\PYG{o}{\PYGZhy{}\PYGZgt{}}\PYG{n}{execute}\PYG{p}{(}\PYG{n}{buffer}\PYG{p}{,} \PYG{n}{circuits}\PYG{p}{);}
\end{Verbatim}
\end{tcolorbox}
\caption{Demonstration of configuring, getting, and executing a remote Accelerator.}
\label{fig:ibm_pig}
\end{figure}

XACC defines a number of \texttt{Compiler} implementations, primary of which are the Quil, OpenQasm, and XASM compilers. All of these implementations leverage the \texttt{Antlr} parser generation library for creating parse-tree listeners that map individual \texttt{Antlr} nodes to XACC IR instructions and composite instructions. The default \texttt{Compiler} in XACC is the XASM compiler (XACC assembly language compiler). This compiler puts forward a grammar specification for a C or C\texttt{++} style low-level assembly language for quantum computing. This language is QASM-like in nature, but inherits the features of the XACC IR static and dynamic instruction set. In this way, XASM provides a dynamic, living language (living in the sense that low-level instructions that can be expressed are in fact XACC Instruction plugins, and can be added/removed at runtime). The XASM grammar specifies instructions with a unique name, a set of bits to operate on, an optional set of gate parameters, and an optional heterogeneous map of input parameters. Moreover, the XASM grammar supports simple \texttt{for} loops that enable programming multiple instructions in a more efficient and familiar manner. These XASM language source strings are parsed via auto-generated \texttt{Antlr} data structures and mapped to XACC IR via an \texttt{Antlr} tree-node visitation event listener implementation. The OpenQasm and Quil compilers are implemented in a similar way, but are based off of vendor-supplied grammar specifications.

XACC defines \texttt{Accelerator} implementations for IBM, Rigetti, D-Wave, IonQ, and a number of back-end simulators (TNQVM \cite{tnqvm-plos-one}, C\texttt{++} local IBM noise-aware simulator, etc). Each of these physical QPU implementations actually subclass a \texttt{RemoteAccelerator} class, which further subclass \texttt{Accelerator}. This remote subtype provides key functionality (that is common across currently available physical QPUs) for remote REST API calls (GET, POST, etc.). The base implementation relies on the Curl for People (CPR) implementation hosted at \cite{cpr}. These \texttt{Accelerators} are further parameterized by pertinent execution metadata such as number of shots, specific back-end, etc. We note that these \texttt{Accelerators} do not delegate to provided Pythonic frameworks (qiskit, pyquil, etc.) but instead make direct calls to the exposed REST API provided by the vendors. An example of getting reference to the IBM \texttt{Accelerator} targeting the Poughkeepsie back-end with 8192 shots is provided in Fig. \ref{fig:ibm_pig}.

\begin{figure}[!t]
\centering
\begin{tcolorbox}[colback=white]
\begin{Verbatim}[commandchars=\\\{\}]
\PYG{n}{\PYGZus{}\PYGZus{}qpu\PYGZus{}\PYGZus{}} \PYG{n+nf}{foo}\PYG{p}{(}\PYG{n}{qbit} \PYG{n}{q}\PYG{p}{,} \PYG{k+kt}{double} \PYG{n}{h}\PYG{p}{,} \PYG{k+kt}{double} \PYG{n}{j}\PYG{p}{)} \PYG{p}{\PYGZob{}}
  \PYG{n}{qmi}\PYG{p}{(}\PYG{n}{q}\PYG{p}{[}\PYG{l+m+mi}{0}\PYG{p}{],}\PYG{n}{q}\PYG{p}{[}\PYG{l+m+mi}{0}\PYG{p}{],} \PYG{n}{h}\PYG{p}{);}
  \PYG{n}{qmi}\PYG{p}{(}\PYG{n}{q}\PYG{p}{[}\PYG{l+m+mi}{1}\PYG{p}{],}\PYG{n}{q}\PYG{p}{[}\PYG{l+m+mi}{1}\PYG{p}{],} \PYG{n}{h}\PYG{p}{);}
  \PYG{n}{qmi}\PYG{p}{(}\PYG{n}{q}\PYG{p}{[}\PYG{l+m+mi}{0}\PYG{p}{],}\PYG{n}{q}\PYG{p}{[}\PYG{l+m+mi}{1}\PYG{p}{],} \PYG{n}{j}\PYG{p}{);}
\PYG{p}{\PYGZcb{}}
\end{Verbatim}
\end{tcolorbox}
\caption{An example quantum kernel for the D-Wave architecture.}
\label{fig:dwave_k}
\end{figure}

XACC provides two implementations of the \texttt{Observable} interface: the \texttt{PauliOperator} and the \texttt{FermionOperator}. The \texttt{PauliOperator} keeps track internally of a mapping of terms, with each term encapsulating a coefficient, a map of qubit sites to string X, Y, Z, and an optional variable string (for parameterized operators). The \texttt{FermionOperator} internally keeps a map of terms, each term containing a coefficient, map of fermionic sites to bool indicating creation or annhilation, and an optional variable string. Both operators expose an API for common algebraic operations on operators (addition, subtraction, multiplication). Both leverage a custom \texttt{Antlr} parser for enabling the \texttt{Observable::fromString} functionality (\texttt{PauliOperators} can be created from strings like \texttt{X0 X1}, \texttt{FermionOperators} can be created from strings like \texttt{0\^{} 1}). \texttt{PauliOperator}  implements \texttt{observe()} to take an unmeasured \texttt{CompositeInstruction}, and for each term the operator contains append appropriate measurement gates, and return a list of measured \texttt{CompositeInstructions}. The \texttt{FermionOperator} implements \texttt{observe()} to first transform itself via an appropriate \texttt{ObservableTransform} (Jordan-Wigner, for instance) and then call and return the result of the \texttt{observe()} call invoked on the transformed \texttt{Observable}.

XACC provides support for the D-Wave quantum computing architecture by implementing the XACC core interfaces for expressing low-level machine instructions that describe biases and couplers for the D-Wave Ising Hamiltonian. Specifically we implement the \texttt{Instruction} interface for D-Wave quantum machine instructions that describe the bias or coupling strength on a given qubit or pair of qubits. XACC defines the \texttt{AnnealingInstruction} sub-type of \texttt{Instruction} which keeps track of two qubit indices and an \texttt{InstructionParameter} parameter for describing the bias or coupler (which can be a concrete double or variable). If the two qubit indices are equal, the instruction describes a bias value, and if they are different, it describes a coupler. Fig.~\ref{fig:dwave_k} provides an example of programming an XACC quantum kernel with these D-Wave instructions. Of course, one could also subclass \texttt{CompositeInstruction} and implement \texttt{expand()} to auto-generate these instructions at runtime.

XACC also defines an \texttt{EmbeddingAlgorithm} interface for the injection of minor graph embedding strategies used in the mapping of these quantum machine instruction representations onto the physical hardware.

\section{Demonstration}
\label{sec:demo}
\begin{figure}[!b]
\centering
\begin{tcolorbox}[colback=white]
\begin{Verbatim}[commandchars=\\\{\}]
\PYG{c+cp}{\PYGZsh{}include} \PYG{c+cpf}{\PYGZdq{}xacc.hpp\PYGZdq{}}
\PYG{k+kt}{int} \PYG{n+nf}{main}\PYG{p}{(}\PYG{k+kt}{int} \PYG{n}{argc}\PYG{p}{,} \PYG{k+kt}{char} \PYG{o}{**}\PYG{n}{argv}\PYG{p}{)} \PYG{p}{\PYGZob{}}
  \PYG{n}{xacc}\PYG{o}{::}\PYG{n}{Initialize}\PYG{p}{(}\PYG{n}{argc}\PYG{p}{,} \PYG{n}{argv}\PYG{p}{);}
  \PYG{c+c1}{// Get reference to the Accelerator}
  \PYG{k}{auto} \PYG{n}{accelerator} \PYG{o}{=}
    \PYG{n}{xacc}\PYG{o}{::}\PYG{n}{getAccelerator}\PYG{p}{(}\PYG{l+s}{\PYGZdq{}tnqvm\PYGZdq{}}\PYG{p}{);}

  \PYG{c+c1}{// Get the IRProvider and create an}
  \PYG{c+c1}{// empty CompositeInstruction}
  \PYG{k}{auto} \PYG{n}{provider} \PYG{o}{=}
    \PYG{n}{xacc}\PYG{o}{::}\PYG{n}{getIRProvider}\PYG{p}{(}\PYG{l+s}{\PYGZdq{}quantum\PYGZdq{}}\PYG{p}{);}
  \PYG{k}{auto} \PYG{n}{program} \PYG{o}{=}
    \PYG{n}{provider}\PYG{o}{\PYGZhy{}\PYGZgt{}}\PYG{n}{createComposite}\PYG{p}{(}\PYG{l+s}{\PYGZdq{}foo\PYGZdq{}}\PYG{p}{,} \PYG{p}{\PYGZob{}}\PYG{l+s}{\PYGZdq{}t\PYGZdq{}}\PYG{p}{\PYGZcb{});}

  \PYG{c+c1}{// Create X, Ry, CX, and Measure gates}
  \PYG{k}{auto} \PYG{n}{x} \PYG{o}{=}
    \PYG{n}{provider}\PYG{o}{\PYGZhy{}\PYGZgt{}}\PYG{n}{createInstruction}\PYG{p}{(}\PYG{l+s}{\PYGZdq{}X\PYGZdq{}}\PYG{p}{,} \PYG{p}{\PYGZob{}}\PYG{l+m+mi}{0}\PYG{p}{\PYGZcb{});}
  \PYG{k}{auto} \PYG{n}{ry} \PYG{o}{=}
    \PYG{n}{provider}\PYG{o}{\PYGZhy{}\PYGZgt{}}\PYG{n}{createInstruction}\PYG{p}{(}\PYG{l+s}{\PYGZdq{}Ry\PYGZdq{}}\PYG{p}{,} \PYG{p}{\PYGZob{}}\PYG{l+m+mi}{1}\PYG{p}{\PYGZcb{},}
                                    \PYG{p}{\PYGZob{}}\PYG{l+s}{\PYGZdq{}t\PYGZdq{}}\PYG{p}{\PYGZcb{});}
  \PYG{k}{auto} \PYG{n}{cx} \PYG{o}{=}
    \PYG{n}{provider}\PYG{o}{\PYGZhy{}\PYGZgt{}}\PYG{n}{createInstruction}\PYG{p}{(}\PYG{l+s}{\PYGZdq{}CNOT\PYGZdq{}}\PYG{p}{,} \PYG{p}{\PYGZob{}}\PYG{l+m+mi}{1}\PYG{p}{,}\PYG{l+m+mi}{0}\PYG{p}{\PYGZcb{});}
  \PYG{k}{auto} \PYG{n}{m0} \PYG{o}{=}
    \PYG{n}{provider}\PYG{o}{\PYGZhy{}\PYGZgt{}}\PYG{n}{createInstruction}\PYG{p}{(}\PYG{l+s}{\PYGZdq{}Measure\PYGZdq{}}\PYG{p}{,}
                                        \PYG{p}{\PYGZob{}}\PYG{l+m+mi}{0}\PYG{p}{\PYGZcb{});}

  \PYG{c+c1}{// Add them to the CompositeInstruction}
  \PYG{n}{program}\PYG{o}{\PYGZhy{}\PYGZgt{}}\PYG{n}{addInstructions}\PYG{p}{(\PYGZob{}}\PYG{n}{x}\PYG{p}{,}\PYG{n}{ry}\PYG{p}{,}\PYG{n}{cx}\PYG{p}{,}\PYG{n}{m0}\PYG{p}{\PYGZcb{});}

  \PYG{c+c1}{// Loop over [\PYGZhy{}pi, pi] and compute \PYGZlt{}Z0\PYGZgt{}}
  \PYG{k}{auto} \PYG{n}{angles} \PYG{o}{=} \PYG{n}{xacc}\PYG{o}{::}\PYG{n}{linspace}\PYG{p}{(}\PYG{o}{\PYGZhy{}}\PYG{n}{pi}\PYG{p}{,} \PYG{n}{pi}\PYG{p}{,} \PYG{l+m+mi}{20}\PYG{p}{);}
  \PYG{k}{for} \PYG{p}{(}\PYG{k}{auto} \PYG{o}{\PYGZam{}}\PYG{n+nl}{a} \PYG{p}{:} \PYG{n}{angles}\PYG{p}{)} \PYG{p}{\PYGZob{}}
    \PYG{k}{auto} \PYG{n}{buffer} \PYG{o}{=} \PYG{n}{xacc}\PYG{o}{::}\PYG{n}{qalloc}\PYG{p}{(}\PYG{l+m+mi}{2}\PYG{p}{);}
    \PYG{k}{auto} \PYG{n}{evaled} \PYG{o}{=} \PYG{n}{program}\PYG{o}{\PYGZhy{}\PYGZgt{}}\PYG{k}{operator}\PYG{p}{()(\PYGZob{}}\PYG{n}{a}\PYG{p}{\PYGZcb{});}
    \PYG{n}{accelerator}\PYG{o}{\PYGZhy{}\PYGZgt{}}\PYG{n}{execute}\PYG{p}{(}\PYG{n}{buffer}\PYG{p}{,} \PYG{n}{evaled}\PYG{p}{);}
    \PYG{n}{std}\PYG{o}{::}\PYG{n}{cout} \PYG{o}{\PYGZlt{}\PYGZlt{}} \PYG{l+s}{\PYGZdq{}\PYGZlt{}Z0\PYGZgt{}(\PYGZdq{}} \PYG{o}{\PYGZlt{}\PYGZlt{}} \PYG{n}{a} \PYG{o}{\PYGZlt{}\PYGZlt{}} \PYG{l+s}{\PYGZdq{}) = \PYGZdq{}}
        \PYG{o}{\PYGZlt{}\PYGZlt{}} \PYG{n}{buffer}\PYG{o}{\PYGZhy{}\PYGZgt{}}\PYG{n}{getExpectationValueZ}\PYG{p}{()}
        \PYG{o}{\PYGZlt{}\PYGZlt{}} \PYG{l+s}{\PYGZdq{}}\PYG{l+s+se}{\PYGZbs{}n}\PYG{l+s}{\PYGZdq{}}\PYG{p}{;}
  \PYG{p}{\PYGZcb{}}

  \PYG{n}{xacc}\PYG{o}{::}\PYG{n}{Finalize}\PYG{p}{();}
  \PYG{k}{return} \PYG{l+m+mi}{0}\PYG{p}{;}
\PYG{p}{\PYGZcb{}}
\end{Verbatim}

\end{tcolorbox}
\caption{Demonstration of programmatically constructing quantum programs as XACC IR. Here we construct a parameterized gate model circuit, and evaluate it at a number of concrete angles for execution on the TNQVM \cite{tnqvm-plos-one} simulation back-end. }
\label{fig:base_api}
\end{figure}

Here we seek to demonstrate the utility of the XACC framework through a few concrete, common use cases. Specifically, we attempt to demonstrate use cases through a variety of XACC abstraction levels, beginning with low-level use cases depicting the base XACC API, and ending with complex hybrid quantum-classical algorithms that leverage custom \texttt{Observables} and \texttt{Algorithms}. The example code shown here is meant to present complete, working programs.

\begin{figure}[!b]
\centering
\begin{tcolorbox}[colback=white]
\begin{Verbatim}[commandchars=\\\{\}]
\PYG{c+cp}{\PYGZsh{}include} \PYG{c+cpf}{\PYGZdq{}xacc.hpp\PYGZdq{}}
\PYG{k+kt}{int} \PYG{n+nf}{main}\PYG{p}{(}\PYG{k+kt}{int} \PYG{n}{argc}\PYG{p}{,} \PYG{k+kt}{char} \PYG{o}{**}\PYG{n}{argv}\PYG{p}{)} \PYG{p}{\PYGZob{}}
  \PYG{n}{xacc}\PYG{o}{::}\PYG{n}{Initialize}\PYG{p}{(}\PYG{n}{argc}\PYG{p}{,} \PYG{n}{argv}\PYG{p}{);}

  \PYG{c+c1}{// Get reference to the Accelerator}
  \PYG{k}{auto} \PYG{n}{accelerator} \PYG{o}{=}
     \PYG{n}{xacc}\PYG{o}{::}\PYG{n}{getAccelerator}\PYG{p}{(}\PYG{l+s}{\PYGZdq{}ibm:ibmq\PYGZus{}valencia\PYGZdq{}}\PYG{p}{);}

  \PYG{c+c1}{// Allocate some qubits}
  \PYG{k}{auto} \PYG{n}{buffer} \PYG{o}{=} \PYG{n}{xacc}\PYG{o}{::}\PYG{n}{qalloc}\PYG{p}{(}\PYG{l+m+mi}{2}\PYG{p}{);}

  \PYG{c+c1}{// Compile}
  \PYG{k}{auto} \PYG{n}{quil} \PYG{o}{=} \PYG{n}{xacc}\PYG{o}{::}\PYG{n}{getCompiler}\PYG{p}{(}\PYG{l+s}{\PYGZdq{}quil\PYGZdq{}}\PYG{p}{);}
  \PYG{k}{auto} \PYG{n}{ir} \PYG{o}{=} \PYG{n}{quil}\PYG{o}{\PYGZhy{}\PYGZgt{}}\PYG{n}{compile}\PYG{p}{(}
    \PYG{l+s+sa}{R}\PYG{l+s}{\PYGZdq{}}\PYG{l+s+dl}{(}\PYG{l+s}{\PYGZus{}\PYGZus{}qpu\PYGZus{}\PYGZus{} void bell(qbit q) \PYGZob{}}
\PYG{l+s}{       H 0}
\PYG{l+s}{       CX 0 1}
\PYG{l+s}{       MEASURE 0 [0]}
\PYG{l+s}{       MEASURE 1 [1]}
\PYG{l+s}{       \PYGZcb{}}\PYG{l+s+dl}{)}\PYG{l+s}{\PYGZdq{}}\PYG{p}{,} \PYG{n}{accelerator}\PYG{p}{);}

  \PYG{c+c1}{// Execute}
  \PYG{n}{accelerator}\PYG{o}{\PYGZhy{}\PYGZgt{}}\PYG{n}{execute}\PYG{p}{(}\PYG{n}{buffer}\PYG{p}{,}
            \PYG{n}{ir}\PYG{o}{\PYGZhy{}\PYGZgt{}}\PYG{n}{getComposites}\PYG{p}{()[}\PYG{l+m+mi}{0}\PYG{p}{]);}

  \PYG{c+c1}{// View results}
  \PYG{n}{buffer}\PYG{o}{\PYGZhy{}\PYGZgt{}}\PYG{n}{print}\PYG{p}{();}

  \PYG{n}{xacc}\PYG{o}{::}\PYG{n}{Finalize}\PYG{p}{();}
  \PYG{k}{return} \PYG{l+m+mi}{0}\PYG{p}{;}
\PYG{p}{\PYGZcb{}}
\end{Verbatim}

\end{tcolorbox}
\caption{Demonstration of creating raw quantum kernel source strings and creating IR via compilation with an appropriate XACC \texttt{Compiler} instance. }
\label{fig:compile_from_qkernel}
\end{figure}

\subsection{Base API}
Fig. \ref{fig:base_api} demonstrates how one might use the lowest level of the XACC API to construct a parameterized quantum program and execute it on a desired back-end. All XACC programs simply include the public XACC header file which provides access to the public XACC API discussed in Section \ref{sec:api}. In order to start requesting services of a given type, programmers must first initialize the framework with the \texttt{xacc::Initialize()} call. This loads all available plugins (implementations of the core XACC interfaces) and parses command line options. Programmers next get reference to the desired \texttt{Accelerator} back-end through the \texttt{getAccelerator()} call. Next, to programmatically construct XACC IR, we get reference to an instance of the \texttt{IRProvider} and use it to create \texttt{Instructions} and \texttt{CompositeInstructions}. All concrete \texttt{Instructions} are added to the \texttt{CompositeInstruction} after creation. We now have a parameterized instance of the XACC IR (parameterized on the variable \texttt{t}, the rotation angle for the \texttt{Ry} instruction). Now, we are free to execute this \texttt{CompositeInstruction} at concrete instances of the variable \texttt{t}. We note that there is no reconstruction of the program or circuit at each \texttt{t}. We simply evaluate this \texttt{CompositeInstruction} at the given variable value and execute that concrete program. Execution is affected by a call to \texttt{Accelerator::execute()} which takes as input the user-provided buffer and the concrete, evaluated program. The \texttt{Accelerator::execute()} implementation is responsible for persisting all execution results to the provided buffer. Afterwards, we are then free to process the results, here we get the expected value in the computational basis. All XACC programs end with a call to \texttt{xacc::Finalize()} which tears down the core service registry and cleans up allocated memory.

\subsection{Kernel Source Compilation and Execution}
We note that the example in Fig. \ref{fig:base_api} is verbose and requires a lot of boilerplate code to express relatively simple concepts. So now we move up to a higher level of XACC abstraction, specifically, compiling quantum kernel source strings via \texttt{Compilers} and the \texttt{xacc::qasm()} call.

As a dual-source approach to quantum-accelerated computing, programmers express quantum code intended for compilation and execution as separate kernel source strings. The code in Fig. \ref{fig:compile_from_qkernel} demonstrates this functionality, specifically in programming a Bell state computation on the remotely hosted IBM Valencia back-end. Programmers begin with the usual \texttt{Initialize} and \texttt{getAccelerator} calls, and instead of programmatically constructing XACC IR, get reference to a specific \texttt{Compiler} implementation to be used in kernel compilation. Programmers compose the quantum kernel code as a string (or string literal) and compile it with the correct \texttt{Compiler} instance, which creates and returns an instance of the IR, ready to be used in back-end execution. This compilation step can optionally take the \texttt{Accelerator} as input, thereby exposing connectivity, noise, or other back-end-specific properties at compile time.

Of course, we can replace this get-compiler, compile, get-IR, workflow with a single call to the \texttt{xacc::qasm()} function, depicted in Fig.~\ref{fig:qasm_generator}.
\begin{figure}[!t]
\centering
\begin{tcolorbox}[colback=white]
\begin{Verbatim}[commandchars=\\\{\}]
\PYG{c+cp}{\PYGZsh{}include} \PYG{c+cpf}{\PYGZdq{}xacc.hpp\PYGZdq{}}
\PYG{k+kt}{int} \PYG{n+nf}{main}\PYG{p}{(}\PYG{k+kt}{int} \PYG{n}{argc}\PYG{p}{,} \PYG{k+kt}{char} \PYG{o}{**}\PYG{n}{argv}\PYG{p}{)} \PYG{p}{\PYGZob{}}
  \PYG{n}{xacc}\PYG{o}{::}\PYG{n}{Initialize}\PYG{p}{(}\PYG{n}{argc}\PYG{p}{,} \PYG{n}{argv}\PYG{p}{);}

  \PYG{c+c1}{// Get the desired Accelerator and Optimizer}
  \PYG{k}{auto} \PYG{n}{qpu} \PYG{o}{=}
    \PYG{n}{xacc}\PYG{o}{::}\PYG{n}{getAccelerator}\PYG{p}{(}\PYG{l+s}{\PYGZdq{}ibm:ibmq\PYGZus{}valencia\PYGZdq{}}\PYG{p}{);}
  \PYG{k}{auto} \PYG{n}{optimizer} \PYG{o}{=} \PYG{n}{xacc}\PYG{o}{::}\PYG{n}{getOptimizer}\PYG{p}{(}\PYG{l+s}{\PYGZdq{}nlopt\PYGZdq{}}\PYG{p}{);}

  \PYG{c+c1}{// Create the N=3 deuteron Hamiltonian}
  \PYG{k}{auto} \PYG{n}{H\PYGZus{}N\PYGZus{}3} \PYG{o}{=} \PYG{n}{xacc}\PYG{o}{::}\PYG{n}{quantum}\PYG{o}{::}\PYG{n}{getObservable}\PYG{p}{(}
     \PYG{l+s}{\PYGZdq{}pauli\PYGZdq{}}\PYG{p}{,}
     \PYG{n}{std}\PYG{o}{::}\PYG{n}{string}\PYG{p}{(}\PYG{l+s}{\PYGZdq{}15.531709 \PYGZhy{} 2.1433 X0X1 \PYGZhy{} \PYGZdq{}}
        \PYG{l+s}{\PYGZdq{}2.1433 Y0Y1 + .21829 Z0 \PYGZhy{} 6.125 Z1 \PYGZhy{} \PYGZdq{}}
        \PYG{l+s}{\PYGZdq{}9.625 Z2 \PYGZhy{} 3.91 X1 X2 \PYGZhy{} 3.91 Y1 Y2\PYGZdq{}}\PYG{p}{));}

  \PYG{c+c1}{// JIT map Quil QASM Ansatz to IR}
  \PYG{n}{xacc}\PYG{o}{::}\PYG{n}{qasm}\PYG{p}{(}\PYG{l+s+sa}{R}\PYG{l+s}{\PYGZdq{}}\PYG{l+s+dl}{(}
\PYG{l+s}{    .compiler xasm}
\PYG{l+s}{    .circuit ansatz}
\PYG{l+s}{    .parameters t0, t1}
\PYG{l+s}{    .qbit q}
\PYG{l+s}{    X(q[0]);}
\PYG{l+s}{    exp\PYGZus{}i\PYGZus{}theta(q, t0,}
\PYG{l+s}{        \PYGZob{}\PYGZob{}\PYGZdq{}pauli\PYGZdq{}, \PYGZdq{}X0 Y1 \PYGZhy{} Y0 X1\PYGZdq{}\PYGZcb{}\PYGZcb{});}
\PYG{l+s}{    exp\PYGZus{}i\PYGZus{}theta(q, t1,}
\PYG{l+s}{        \PYGZob{}\PYGZob{}\PYGZdq{}pauli\PYGZdq{}, \PYGZdq{}X0 Z1 Y2 \PYGZhy{} X2 Z1 Y0\PYGZdq{}\PYGZcb{}\PYGZcb{});}
\PYG{l+s}{    }\PYG{l+s+dl}{)}\PYG{l+s}{\PYGZdq{}}\PYG{p}{);}
  \PYG{k}{auto} \PYG{n}{ansatz} \PYG{o}{=} \PYG{n}{xacc}\PYG{o}{::}\PYG{n}{getCompiled}\PYG{p}{(}\PYG{l+s}{\PYGZdq{}ansatz\PYGZdq{}}\PYG{p}{);}

  \PYG{c+c1}{// Get the VQE Algorithm and initialize it}
  \PYG{k}{auto} \PYG{n}{vqe} \PYG{o}{=} \PYG{n}{xacc}\PYG{o}{::}\PYG{n}{getAlgorithm}\PYG{p}{(}\PYG{l+s}{\PYGZdq{}vqe\PYGZdq{}}\PYG{p}{);}
  \PYG{n}{vqe}\PYG{o}{\PYGZhy{}\PYGZgt{}}\PYG{n}{initialize}\PYG{p}{(\PYGZob{}}
        \PYG{n}{std}\PYG{o}{::}\PYG{n}{make\PYGZus{}pair}\PYG{p}{(}\PYG{l+s}{\PYGZdq{}ansatz\PYGZdq{}}\PYG{p}{,} \PYG{n}{ansatz}\PYG{p}{),}
        \PYG{n}{std}\PYG{o}{::}\PYG{n}{make\PYGZus{}pair}\PYG{p}{(}\PYG{l+s}{\PYGZdq{}observable\PYGZdq{}}\PYG{p}{,} \PYG{n}{H\PYGZus{}N\PYGZus{}3}\PYG{p}{),}
        \PYG{n}{std}\PYG{o}{::}\PYG{n}{make\PYGZus{}pair}\PYG{p}{(}\PYG{l+s}{\PYGZdq{}accelerator\PYGZdq{}}\PYG{p}{,} \PYG{n}{qpu}\PYG{p}{),}
        \PYG{n}{std}\PYG{o}{::}\PYG{n}{make\PYGZus{}pair}\PYG{p}{(}\PYG{l+s}{\PYGZdq{}optimizer\PYGZdq{}}\PYG{p}{,} \PYG{n}{optimizer}\PYG{p}{)}
        \PYG{p}{\PYGZcb{});}

  \PYG{c+c1}{// Allocate some qubits and execute}
  \PYG{k}{auto} \PYG{n}{buffer} \PYG{o}{=} \PYG{n}{xacc}\PYG{o}{::}\PYG{n}{qalloc}\PYG{p}{(}\PYG{l+m+mi}{3}\PYG{p}{);}
  \PYG{n}{vqe}\PYG{o}{\PYGZhy{}\PYGZgt{}}\PYG{n}{execute}\PYG{p}{(}\PYG{n}{buffer}\PYG{p}{);}

  \PYG{c+c1}{// Get the result}
  \PYG{k}{auto} \PYG{n}{energy} \PYG{o}{=}
        \PYG{p}{(}\PYG{o}{*}\PYG{n}{buffer}\PYG{p}{)[}\PYG{l+s}{\PYGZdq{}opt\PYGZhy{}val\PYGZdq{}}\PYG{p}{].}\PYG{n}{as}\PYG{o}{\PYGZlt{}}\PYG{k+kt}{double}\PYG{o}{\PYGZgt{}}\PYG{p}{();}

  \PYG{n}{xacc}\PYG{o}{::}\PYG{n}{Finalize}\PYG{p}{();}
  \PYG{k}{return} \PYG{l+m+mi}{0}\PYG{p}{;}
\PYG{p}{\PYGZcb{}}
\end{Verbatim}

\end{tcolorbox}
\caption{Demonstration of a complex quantum-classical algorithm using \texttt{Observable}, \texttt{Optimizer}, and \texttt{Algorithm}. Specifically, this code snippet executes the variational quantum eigensolver to compute the ground state energy of the three qubit deuteron Hamiltonian in \cite{Dumitrescu2018}.}
\label{fig:deuteronH3}
\end{figure}

\subsection{Hybrid Quantum-Classical Algorithms}
Here we demonstrate leveraging the highest-level utilities in XACC to execute a typical quantum-classical algorithm. Specifically, we consider two common applications: (1) computing the binding energy of deuteron using the $N=3$ Hamiltonian from \cite{Dumitrescu2018} using the variational quantum eigensolver (VQE), and (2) training a quantum circuit to learn a provided target probability distribution.

First we consider the application from nuclear physics, with XACC code demonstrated in Fig.~\ref{fig:deuteronH3}. This example demonstrates the utility of the various high-level abstractions discussed in the preceding discussion, specifically the \texttt{Observable}, \texttt{Optimizer}, and \texttt{Algorithm}.
Specifically we seek to compute the ground state energy of
\begin{align}
H_3 &= 15.531709 I +0.218291Z_0 -6.125 Z_1 \nonumber\\
& -2.143304 \left(X_0 X_1 + Y_0Y_1\right) - 9.625 Z_2 \nonumber\\
& -3.913119\left(X_1 X_2 +Y_1Y_2\right) \label{H3} .
\end{align}
with the variational ansatz given by
\begin{equation}
    U(\eta, \theta) \equiv e^{-i{\eta\over 2}\left(X_0 Y_1 - X_1 Y_0 \right)} e^{-i{\theta\over 2}\left(X_0 Z_1 Y_2 - X_2 Z_1 Y_0\right)}.
    \label{eq:u}
\end{equation}
\begin{figure}[!t]
\centering
\begin{tcolorbox}[colback=white]
\begin{Verbatim}[commandchars=\\\{\}]
\PYG{c+cp}{\PYGZsh{}include} \PYG{c+cpf}{\PYGZdq{}xacc.hpp\PYGZdq{}}
\PYG{k+kt}{int} \PYG{n+nf}{main}\PYG{p}{(}\PYG{k+kt}{int} \PYG{n}{argc}\PYG{p}{,} \PYG{k+kt}{char} \PYG{o}{**}\PYG{n}{argv}\PYG{p}{)} \PYG{p}{\PYGZob{}}
  \PYG{n}{xacc}\PYG{o}{::}\PYG{n}{Initialize}\PYG{p}{(}\PYG{n}{argc}\PYG{p}{,} \PYG{n}{argv}\PYG{p}{);}

  \PYG{c+c1}{// Get reference to the Accelerator}
  \PYG{k}{auto} \PYG{n}{qpu} \PYG{o}{=}
      \PYG{n}{xacc}\PYG{o}{::}\PYG{n}{getAccelerator}\PYG{p}{(}\PYG{l+s}{\PYGZdq{}qcs:Aspen\PYGZhy{}4\PYGZhy{}2Q\PYGZhy{}A\PYGZdq{}}\PYG{p}{);}

  \PYG{k}{auto} \PYG{n}{opt} \PYG{o}{=}
            \PYG{n}{xacc}\PYG{o}{::}\PYG{n}{getOptimizer}\PYG{p}{(}\PYG{l+s}{\PYGZdq{}mlpack\PYGZdq{}}\PYG{p}{);}
  \PYG{n}{xacc}\PYG{o}{::}\PYG{n}{qasm}\PYG{p}{(}\PYG{l+s+sa}{R}\PYG{l+s}{\PYGZdq{}}\PYG{l+s+dl}{(}
\PYG{l+s}{.compiler xasm}
\PYG{l+s}{.circuit ansatz}
\PYG{l+s}{.parameters x}
\PYG{l+s}{.qbit q}
\PYG{l+s}{U(q[0], x[0], \PYGZhy{}pi/2, pi/2 );}
\PYG{l+s}{U(q[0], 0, 0, x[1]);}
\PYG{l+s}{U(q[1], x[2], \PYGZhy{}pi/2, pi/2);}
\PYG{l+s}{U(q[1], 0, 0, x[3]);}
\PYG{l+s}{CNOT(q[0], q[1]);}
\PYG{l+s}{U(q[0], 0, 0, x[4]);}
\PYG{l+s}{U(q[0], x[5], \PYGZhy{}pi/2, pi/2);}
\PYG{l+s}{U(q[1], 0, 0, x[6]);}
\PYG{l+s}{U(q[1], x[7], \PYGZhy{}pi/2, pi/2);}
\PYG{l+s+dl}{)}\PYG{l+s}{\PYGZdq{}}\PYG{p}{);}
  \PYG{k}{auto} \PYG{n}{ansatz} \PYG{o}{=} \PYG{n}{xacc}\PYG{o}{::}\PYG{n}{getCompiled}\PYG{p}{(}\PYG{l+s}{\PYGZdq{}ansatz\PYGZdq{}}\PYG{p}{);}

  \PYG{n}{std}\PYG{o}{::}\PYG{n}{vector}\PYG{o}{\PYGZlt{}}\PYG{k+kt}{double}\PYG{o}{\PYGZgt{}} \PYG{n}{target\PYGZus{}distribution}
                    \PYG{p}{\PYGZob{}}\PYG{l+m+mf}{.5}\PYG{p}{,} \PYG{l+m+mf}{.5}\PYG{p}{,} \PYG{l+m+mf}{.5}\PYG{p}{,} \PYG{l+m+mf}{.5}\PYG{p}{\PYGZcb{};}

  \PYG{k}{auto} \PYG{n}{ddcl} \PYG{o}{=} \PYG{n}{xacc}\PYG{o}{::}\PYG{n}{getAlgorithm}\PYG{p}{(}\PYG{l+s}{\PYGZdq{}ddcl\PYGZdq{}}\PYG{p}{);}
  \PYG{n}{ddcl}\PYG{o}{\PYGZhy{}\PYGZgt{}}\PYG{n}{initialize}\PYG{p}{(\PYGZob{}}
        \PYG{n}{std}\PYG{o}{::}\PYG{n}{make\PYGZus{}pair}\PYG{p}{(}\PYG{l+s}{\PYGZdq{}ansatz\PYGZdq{}}\PYG{p}{,} \PYG{n}{ansatz}\PYG{p}{),}
        \PYG{n}{std}\PYG{o}{::}\PYG{n}{make\PYGZus{}pair}\PYG{p}{(}\PYG{l+s}{\PYGZdq{}target\PYGZus{}dist\PYGZdq{}}\PYG{p}{,}
                \PYG{n}{target\PYGZus{}distribution}\PYG{p}{),}
        \PYG{n}{std}\PYG{o}{::}\PYG{n}{make\PYGZus{}pair}\PYG{p}{(}\PYG{l+s}{\PYGZdq{}accelerator\PYGZdq{}}\PYG{p}{,} \PYG{n}{qpu}\PYG{p}{),}
        \PYG{n}{std}\PYG{o}{::}\PYG{n}{make\PYGZus{}pair}\PYG{p}{(}\PYG{l+s}{\PYGZdq{}loss\PYGZdq{}}\PYG{p}{,} \PYG{l+s}{\PYGZdq{}js\PYGZdq{}}\PYG{p}{),}
        \PYG{n}{std}\PYG{o}{::}\PYG{n}{make\PYGZus{}pair}\PYG{p}{(}\PYG{l+s}{\PYGZdq{}gradient\PYGZdq{}}\PYG{p}{,}
                \PYG{l+s}{\PYGZdq{}js\PYGZhy{}parameter\PYGZhy{}shift\PYGZdq{}}\PYG{p}{),}
        \PYG{n}{std}\PYG{o}{::}\PYG{n}{make\PYGZus{}pair}\PYG{p}{(}\PYG{l+s}{\PYGZdq{}optimizer\PYGZdq{}}\PYG{p}{,} \PYG{n}{opt}\PYG{p}{)\PYGZcb{});}

  \PYG{c+c1}{// Allocate some qubits and execute}
  \PYG{k}{auto} \PYG{n}{buffer} \PYG{o}{=} \PYG{n}{xacc}\PYG{o}{::}\PYG{n}{qalloc}\PYG{p}{(}\PYG{l+m+mi}{2}\PYG{p}{);}
  \PYG{n}{ddcl}\PYG{o}{\PYGZhy{}\PYGZgt{}}\PYG{n}{execute}\PYG{p}{(}\PYG{n}{buffer}\PYG{p}{);}

  \PYG{c+c1}{// Print the result}
  \PYG{n}{std}\PYG{o}{::}\PYG{n}{cout} \PYG{o}{\PYGZlt{}\PYGZlt{}} \PYG{l+s}{\PYGZdq{}Loss: \PYGZdq{}} \PYG{o}{\PYGZlt{}\PYGZlt{}}
        \PYG{p}{(}\PYG{o}{*}\PYG{n}{buffer}\PYG{p}{)[}\PYG{l+s}{\PYGZdq{}opt\PYGZhy{}val\PYGZdq{}}\PYG{p}{].}\PYG{n}{as}\PYG{o}{\PYGZlt{}}\PYG{k+kt}{double}\PYG{o}{\PYGZgt{}}\PYG{p}{()}
            \PYG{o}{\PYGZlt{}\PYGZlt{}} \PYG{l+s}{\PYGZdq{}}\PYG{l+s+se}{\PYGZbs{}n}\PYG{l+s}{\PYGZdq{}}\PYG{p}{;}
\PYG{p}{\PYGZcb{}}
\end{Verbatim}

\end{tcolorbox}
\caption{Demonstration of using XACC and the \texttt{Algorithm} interface for quantum-enhanced machine learning tasks.}
\label{fig:ddcl}
\end{figure}

The code snippet begins with the usual initialization and getting reference to the desired \texttt{Accelerator}. Next, we create an \texttt{Observable} that models the three qubit Hamiltonian. Note specifically how we are able to express this Hamiltonian as it would be written in text. Next we get reference to the default \texttt{NLOpt} \texttt{Optimizer} for our VQE execution. Next, we program and compile the VQE state preparation circuit via the \texttt{xacc::qasm()} call, and we leverage the dynamic runtime \texttt{exp\_i\_theta} \texttt{CompositeInstruction}. This too can be expressed very simply, with a straightforward translation from the equation specified in Eq.~\ref{eq:u}. The dynamic exponentiation instruction will be just-in-time expanded to the appropriate circuit decomposition. Next, we get reference to the VQE \texttt{Algorithm}, which must be initialized with a valid \texttt{ansatz}, \texttt{observable}, \texttt{optimizer}, and \texttt{accelerator}. We then allocate three qubits via the \texttt{xacc::qalloc()} call, and execute the \texttt{Algorithm} with the allocated buffer. This call kicks off the VQE algorithm, and the \texttt{Optimizer} will iteratively converge to the optimal energy over the two circuit parameters \texttt{t0,t1}. The results of the computation can be retrieved from the \texttt{AcceleratorBuffer}, with execution results from every iteration and for every measured quantum program persisted to the provided \texttt{AcceleratorBuffer} as children \texttt{AcceleratorBuffers}.

Next we turn to an application of quantum computation to classical machine learning. Our goal is to train a parameterized quantum circuit to reproduce a target probability distribution \cite{hamilton}. For this data-driven circuit learning (DDCL) process, XACC provides an \texttt{Algorithm} implementation called \texttt{DDCL}. This implementation takes as input the target probability distribution, a \texttt{CompositeInstruction} representing the parameterized circuit, the back-end \texttt{Accelerator} to run on, the \texttt{Optimizer} used for the learning process, and loss function and gradient strategies. With this input, \texttt{DDCL} orchestrates the workflow necessary for to find the optimal circuit parameters that lead to a probability distribution that mimics the given target distribution. Fig.~\ref{fig:ddcl} demonstrates how one would use this \texttt{Algorithm}. The programmer starts by getting reference to the desired \texttt{Accelerator} and \texttt{Optimizer}, here a 2-qubit lattice on the Rigetti QCS platform and the MLPack optimizer \cite{mlpack2018}, respectively. Next we compile an XASM source code with 8 parameters using general, three-parameter IBM \texttt{U} rotation gates and a CNOT. Note the use of IBM (\texttt{U}, \texttt{CNOT}) instructions for Rigetti (\texttt{RZ}, \texttt{CZ}), XACC handles this cross-compilation via subsequent transformations of the IR. We define a target probability distribution and get reference to the \texttt{DDCL} algorithm. We initialize the algorithm with the required input data, specifically requesting that the algorithm compute the Jansen-Shannon divergence of the circuit and target distributions. Moreover, we specify a gradient computation strategy that evaluates the parameterized circuit at $x+\pi/2$ for all $x$. Finally, we allocate a two qubit buffer and execute. Upon completion, the buffer provides all pertinent execution information, as well as any algorithm-specific results.

\section{Discussion}
\label{sec:disc}
We have presented a system-level framework for heterogeneous quantum-classical computing. Our goal with this work is to provide a future-proof approach for enabling software technologies targeting quantum-accelerated heterogeneous computing architectures. XACC provides a modular and extensible approach to hardware-agnostic quantum computation that can serve as a foundation for a variety of software tools, compilers, and benchmarks. We have demonstrated a layered architecture that promotes a dual-source, co-processor programming model that enables quantum code to be expressed as high-level kernels. Our approach provides extension points for mapping these kernel expressions to a core, polymorphic intermediate representation, which is amenable to low-level quantum program optimizations, reductions, and analysis. Our extensible back-end infrastructure supports both gate model and annealing quantum computing technologies, and currently supports processors and simulators from IBM, Rigetti, D-Wave, and IonQ. As quantum processor technologies improve, our approach can serve as the foundation of a comprehensive integration framework for all aspects of the quantum programming software stack. This framework can serve as a means for tighter integration models that promote performant workflows leveraging quantum information processing.

\section*{Acknowledgements}
\label{}
This work has been supported by the Laboratory Directed Research and Development Program of Oak Ridge National Laboratory, the US Department of Energy (DOE) Office of Science Advanced Scientific Computing Research (ASCR) Early Career Research Award, and the DOE Office of Science ASCR  Quantum Computing Application Teams and Quantum Testbed Pathfinder programs, under field work proposal numbers ERKJ347 and ERKJ335, and DOE Office of Science High Energy Physics QuantISED project ERKAP61. This work was also supported by the ORNL Undergraduate Research Participation Program, which is sponsored by ORNL and administered jointly by ORNL and the Oak Ridge Institute for Science and Education (ORISE). ORNL is managed by UT-Battelle, LLC, for the US Department of Energy under contract no. DE-AC05-00OR22725. ORISE is managed by Oak Ridge Associated Universities for the US Department of Energy under contract no. DE-AC05-00OR22750. The US government retains and the publisher, by accepting the article for publication, acknowledges that the US government retains a nonexclusive, paid-up, irrevocable, worldwide license to publish or reproduce the published form of this manuscript, or allow others to do so, for US government purposes. DOE will provide public access to these results of federally sponsored research in accordance with the DOE Public Access Plan.

\bibliography{main}

\end{document}